\documentclass[11pt]{article}
\usepackage{amssymb,amsfonts,latexsym,stmaryrd}
\usepackage[PostScript=dvips]{diagrams}
\usepackage{proof}
\usepackage{QED}
\usepackage{pstricks,pst-node,pst-tree}
\usepackage{epsfig}

\newtheorem{theorem}{Theorem}

\newtheorem{example}[theorem]{Example}
\newtheorem{proposition}[theorem]{Proposition}
\newtheorem{lemma}[theorem]{Lemma}

\newtheorem{definition}[theorem]{Definition}

\newcommand{\lrarr}{\longrightarrow}
\newcommand{\rarr}{\rightarrow}

\newcommand{\CC}{\mathcal{C}}
\newcommand{\DD}{\mathcal{D}}
\newcommand{\eqdef}{:=}
\newcommand{\op}[1]{#1^{\mathsf{op}}}
\newcommand{\CCop}{\op{\CC}}
\newcommand{\ie}{\textit{i.e.}\ }
\newcommand{\sdot}{\bullet}
\newcommand{\id}[1]{\mathsf{id}_{#1}}
\newcommand{\II}{I}
\newcommand{\natarrow}{\stackrel{.}{\lrarr}}

\newcommand{\Obj}{\bullet}
\newcommand{\HH}{\mathcal{H}}
\newcommand{\ket}[1]{{|} #1\rangle}
\renewcommand{\Tr}{\mathsf{Tr}}
\newcommand{\coname}[1]{\llcorner #1 \lrcorner}

\newcommand{\Ax}{\someQWE{Id}}
\newcommand{\someQWE}[1]{\mbox{\small$\mathsf{#1}$}}
\newcommand{\someQWEqwe}[2]{\someQWE{#1\,#2}}
\newcommand{\AndI}{\someQWEqwe{\wedge}{I}}
\newcommand{\AndEl}{\someQWEqwe{\wedge}{E_1}}
\newcommand{\AndEr}{\someQWEqwe{\wedge}{E_2}}
\newcommand{\ImpI}{\someQWEqwe{\mathnormal\supset}{I}}
\newcommand{\ImpE}{\someQWEqwe{\mathnormal\supset}{E}}
\newcommand{\App}[1][]{\mathsf{ev}_{#1}}

\newcommand{\iso}{\cong}
\newcommand{\Rel}{\textbf{Rel}}
\newcommand{\Vect}{\textbf{Vect}}

\newcommand{\Set}{\textbf{Set}}
\newcommand{\FdHilb}{\mathbf{FdHilb}}
\newcommand{\ang}[1]{\langle#1\rangle}
\newcommand{\Complex}{\mathbb{C}}

\newcommand{\dd}{\llcorner}
\newcommand{\ddd}{\lrcorner}
\newcommand{\uu}{\ulcorner} 
\newcommand{\uuu}{\urcorner}

\newcommand{\vsa}{\vspace{.1in}}

\newcommand{\vsn}{\vspace{-.1in}}

\newarrow{LRTo}<--->
\newarrow{inc}C--->
\newarrow{Dotline}.....
\newarrow{eq}=====
\newarrow{Eq}=====
\newarrow{Dash}....{>}
\newarrow{Toto}<--->
\newarrow{Is}===== 
\newarrow{Dotsto}....>
\newarrow{Dots}.....

\title{No-Cloning In Categorical Quantum Mechanics}
\author{Samson Abramsky}
\date{}

\begin{document}

\maketitle

\section{Introduction}
The No-Cloning theorem \cite{Dieks,WZ} is a basic limitative result for quantum mechanics, with particular significance for quantum information. It says that there is no unitary operation which makes perfect copies of an unknown (pure) quantum state. A stronger form of this result is the No-Broadcasting theorem  \cite{Broadcast}, which applies to mixed states. There is also a No-Deleting theorem \cite{Pati}.

Recently, the author and Bob Coecke have introduced a categorical formulation of Quantum Mechanics \cite{AC2,AC3,AC4},
as a basis for a more structural, high-level approach to quantum information and computation. This has been elaborated by ourselves, our colleagues, and other workers in the field \cite{Abr0,Abr1,Abr2,AD,CPav,CD,Selinger,Vicary}, and has been shown to yield an effective and illuminating treatment of a wide range of topics in quantum information.
Diagrammatic calculi for tensor categories \cite{JoyalStreet,Turaev}, suitably extended to incorporate the various additional structures which have been used to reflect fundamental features of quantum mechanics, play an important r\^ole, both as an intuitive and vivid visual presentation of the formalism, and as an effective calculational device.

It is clear that such a novel reformulation of the mathematical formalism of quantum mechanics, a subject more or less set in stone since Von Neumann's classic treatise \cite{vN}, has the potential to yield new insights into the foundations of quantum mechanics. In the present paper, we shall use it to open up a novel perspective on No-Cloning. What we shall find, quite unexpectedly, is a link to some fundamental issues in logic, computation, and the foundations of mathematics.
A striking feature of our results is that they are visibly in the same genre as a well-known result by Joyal in categorical logic \cite{LS} showing that a `Boolean cartesian closed category' trivializes, which provides a major road-block to the computational interpretation of classical logic. In fact, they strengthen Joyal's result, insofar as the assumption of a full categorical product (diagonals \emph{and} projections) in the presence of  a classical duality is weakened. This shows a heretofore unsuspected connection between limitative results in proof theory and No-Go theorems in quantum mechanics.

The further contents of the paper are as follows:
\begin{itemize}
\item In the next section, we shall briefly review the three-way link between logic, computation and categories, and recall Joyal's lemma. 
\item In section~3, we shall review the categorical approach to quantum mechanics.
\item Our main results are in section~4, where we prove our  limitative result, which shows the incompatibility of structural features corresponding to quantum entanglement (essentially, the existence of Bell states enabling teleportation) with the existence of a `natural' (in the categorical sense, corresponding essentially to \emph{basis-independent}) copying operation. This result is mathematically robust, since it is proved in a very general context, and has a topological content which is clearly revealed by a diagrammatic proof. At the same time it is delicately poised, since \emph{non-natural}, basis-dependent copying operations do in fact play a key r\^ole in the categorical formulation of quantum notions of measurement. We discuss this context, and the conceptual reading of the results.
\item We conclude with some discussion of extensions of the results, further directions, and open problems.
\end{itemize}

\section{Categories, Logic and Computational Content: Joyal's Lemma}

Categorical logic \cite{LS} and the Curry-Howard correspondence in Proof Theory \cite{CH} give us a beautiful three-way correspondence:

\begin{diagram}
\mbox{Logic} & & \rLRTo & & \mbox{Computation} \\
& \luLRTo & & \ruLRTo & \\
& & \mbox{Categories} & &
\end{diagram}

More particularly, we have as a paradigmatic example:
\begin{diagram}
\mbox{Intuitionistic Logic} & & \rLRTo & & \mbox{$\lambda$-calculus} \\
& \luLRTo & & \ruLRTo & \\
& & \mbox{Cartesian Closed Categories} & &
\end{diagram}
Here we are focussing on the fragment of intuitionistic logic containing conjunction and implication, and the simply-typed $\lambda$-calculus with product types.

We shall assume familiarity with basic notions of category theory \cite{Mac,LS2}. Recall that a cartesian closed category is a category with a terminal object, binary products and exponentials. The basic cartesian closed adjunction is
\[ \CC(A \times B, C) \cong \CC(A, B \Rightarrow C) \, . \]
More explicitly, a category $\CC$ with finite products \emph{has exponentials} if for all objects $A$ and
$B$ of $\CC$ there is a couniversal arrow from $- \times A$ to $B$,
\textit{i.e.} an object $A \Rightarrow B$ of $\CC$ and a  morphism  
\[ \App[A,B] :
(A \Rightarrow B) \times A \longrightarrow B \]
with the couniversal property: for every $g : C \times A
\longrightarrow B$, there is a unique morphism $\Lambda (g) : C
\longrightarrow A \Rightarrow B$ such that 
\[ \begin{diagram}
A \Rightarrow B \\
\uDashto^{\Lambda (g)} \\
C \\
\end{diagram}
\qquad \qquad
\begin{diagram}
(A \Rightarrow B) \times A  & \rTo^{\App[A,B]}  & B \\
\uDashto^{\Lambda (g) \times \id{A}}  & \ruTo_g & \\
C \times A & & \\
\end{diagram}
\]

\noindent The correspondence between the intuitionistic logic of conjunction and implication and cartesian closed categories is summarized in the following table:
\begin{center}
\renewcommand{\arraystretch}{0.5}\fbox{$\begin{array}{c||@{\;\;}c@{\;\;}|@{\;\;}c@{\;\;}} &&\\
\textbf{Axiom} & \infer[\Ax]{\Gamma,A\vdash A}{} & \infer{\pi_{2}:\Gamma\times A\longrightarrow A}{}
\\&&\\\hline&&\\
\textbf{Conjunction} & \infer[\AndI]{\Gamma\vdash A\wedge B}{\Gamma\vdash A\qquad\Gamma\vdash B} & \infer{\langle f,g\rangle:\Gamma\longrightarrow A
\times B}{f : \Gamma \longrightarrow A \qquad g : \Gamma \longrightarrow B}
\\&&\\
& \infer[\AndEl]{\Gamma \vdash A}{\Gamma \vdash A \wedge B} & \infer{\pi_{1} \circ f : \Gamma \longrightarrow A}{f : \Gamma \longrightarrow A \times B}
\\&&\\
& \infer[\AndEr]{\Gamma \vdash B}{\Gamma \vdash A \wedge B} & \infer{\pi_{2}\circ f : \Gamma \longrightarrow B}{f : \Gamma \longrightarrow A \times B}
\\&&\\\hline&&\\
\textbf{Implication} & \infer[\ImpI]{\Gamma\vdash A\supset B}{\Gamma,A\vdash B} & \infer{\Lambda(f):\Gamma\longrightarrow(A\Rightarrow B)}{f:\Gamma
\times A\longrightarrow B} \\&&\\
& \infer[\ImpE]{\Gamma\vdash B}{\Gamma\vdash A\supset B \quad\Gamma\vdash A} & \infer{\App[A,B]\circ\langle f,g\rangle:\Gamma\longrightarrow B}{f :
\Gamma \longrightarrow (A \Rightarrow B) \quad g : \Gamma \rarr A}\\&&\\
\end{array}$}
\end{center}

\subsection{Joyal's Lemma}
\label{JLsec}

It is a very natural idea to seek to extend the correspondence shown above to the case of \emph{classical logic}. Joyal's lemma shows that there is a fundamental impediment to doing so.\footnote{It is customary to refer to this result as Joyal's lemma, although, apparently, he never published it. The usual reference is \cite{LS}, who attribute the result to Joyal, but follow the proof given by Freyd \cite{Freyd}. Our statement and proof are somewhat different to those in \cite{LS}.}

The natural extension of the notion of cartesian closed category, which corresponds to the \emph{intuitionistic logic} of conjunction and implication, to the classical case  is to introduce a suitable notion of classical negation. We recall that it is customary in intuitionistic logic to \emph{define} the negation by
\[ \neg A \eqdef A \supset \bot \]
where $\bot$ is the \emph{falsum}. The characteristic property of the falsum is that it implies every proposition. In categorical terms, this translates into the notion of an initial object.
Note that for any fixed object $B$ in a cartesian closed category, there is a well-defined contravariant functor
\[ \CC \lrarr \op{\CC} :: A \mapsto (A \Rightarrow B) \, . \]
This will always satisfy the properties corresponding to negation in minimal logic, and if $B = \bot$ is the initial object in $\CC$, then it will satisfy the laws of intuitionistic negation. In particular, there is a canonical arrow
\[ A \lrarr (A \Rightarrow \bot) \Rightarrow \bot \]
which is just the curried form of the evaluation morphism. This  corresponds to the valid intuitionistic principle $A \supset \neg \neg A$.
What else is needed in order to obtain classical logic? As is well known, the missing principle is that of \emph{proof by contradiction}: the converse implication $\neg \neg A \supset A$.

This leads us to the following notion. A \emph{dualizing object} $\bot$ in a closed category is one for which the canonical arrow
\[ A \lrarr (A \Rightarrow \bot) \Rightarrow \bot \]
is an isomorphism for all $A$.

We can now state Joyal's lemma:
\begin{proposition}[Joyal's Lemma]
Any cartesian closed category with a dualizing object is a preorder (hence \emph{trivial} as a semantics for proofs or computational processes).
\end{proposition}

\begin{proof}
Note firstly that, if $\bot$ is dualizing, the induced negation functor $\CC \lrarr \CCop$ is a \emph{contravariant equivalence} $\CC \simeq \CCop$. Since $(\top \Rightarrow A) \cong A$ where $\top$ is the terminal object, it follows that $\bot$ is the dual of $\top$, and hence initial.
So it suffices to prove Joyal's lemma under the assumption that the dualizing object is initial.

We assume that $\bot$ is a dualizing initial object in a cartesian closed category $\CC$. We write $\iota_{C} : \bot \rarr C$ for the unique arrow given by initiality.
Note that  $\CC(A \times \bot, A \times \bot) \cong \CC(\bot, A \Rightarrow (A \times \bot))$, which is a singleton by initiality.
It follows that $\iota_{A \times \bot} \circ \pi_{2} = \id{A \times \bot}$, while $\pi_{2}  \circ \iota_{A \times \bot} = \id{\bot}$ by initiality.
Hence $A \times \bot \cong \bot$.\footnote{A slicker proof simply notes that $A \times (-)$ is a left adjoint by cartesian closure, and hence preserves all colimits, in particular initial objects.}

Now 
\begin{equation}
\label{JLeq}
\CC(A, B) \cong \CC(B \Rightarrow \bot, A \Rightarrow \bot) \cong \CC((B \Rightarrow \bot) \times A, \bot) . 
\end{equation}
Given any $h, k : C \lrarr \bot$, note that
\[ h = \pi_{1} \circ \langle h, k \rangle , \qquad k = \pi_{2} \circ \langle h, k \rangle  . \]
But $\bot \times \bot \cong \bot$, hence by initiality $\pi_{1} = \pi_{2}$, and so $h = k$, which by (\ref{JLeq}) implies that $f = g$ for $f, g : A \lrarr B$.
\end{proof}

\subsection{Linearity and Classicality}
However, we know from Linear Logic that there is no impediment to having a closed structure with a dualizing object, \emph{provided} we weaken our assumption on the underlying context-building structure, from \emph{cartesian} $\times$ to \emph{monoidal} $\otimes$.

Then we get a wealth of examples of \emph{$*$-autonomous categories} \cite{Barr}, which stand to Multiplicative Linear Logic as cartesian closed categories do to Intuitionistic Logic \cite{Seely}.

Joyal's lemma can thus be stated in the following equivalent form.

\begin{proposition}
A $*$-autonomous category in which the monoidal structure is cartesian is a preorder.
\end{proposition}

Essentially, a cartesian structure is a monoidal structure plus natural diagonals, and with the tensor unit a terminal object, \ie  \emph{plus cloning and deleting}!

\section{Categorical Quantum Mechanics}

In this section, we shall provide a brief review of the structures used in categorical quantum mechanics, their graphical representation, and how these structures are used in formalizing some key features of quantum mechanics. Further details can be found elsewhere \cite{AC4,Abr1,Selinger}.

\subsection{Symmetric Monoidal Categories}
We recall that a  \emph{monoidal category} is a structure $(\CC , \otimes , I, a, l, r)$ where:
\begin{itemize}
\item $\CC$ is a category,
\item $\otimes : \CC \times \CC \rarr\CC$ is a functor (\emph{tensor}),
\item $I$ is a distinguished object of $\CC$ (\emph{unit}),
\item $a$, $l$, $r$ are natural isomorphisms (\emph{structural isos}) with components:
\[ a_{A, B, C} : A \otimes (B \otimes C) \iso (A \otimes B) \otimes C \]
\[ l_A : I \otimes A \iso A \qquad \quad r_A : A \otimes I \iso A \]
\end{itemize}
such that certain diagrams commute, which ensure \emph{coherence} \cite{Mac}, described by the slogan:
\begin{center}
\fbox{All diagrams only involving $a$, $l$ and $r$ must commute.}
\end{center}
Examples:
\begin{itemize}
\item Both products and coproducts give rise to monoidal structures\,---\,which are the common denominator between them.
(But in addition, products have \emph{diagonals} and \emph{projections}, and coproducts have \emph{codiagonals} and \emph{injections}.)
\item $(\mathbb{N}, {\leqslant} , + , 0)$ is a monoidal category.
\item $\Rel$, the category of sets and relations, with cartesian  product (which is \emph{not} the categorical product).
\item $\Vect_k$ with the standard tensor product.
\end{itemize}
Let us examine the example of $\Rel$ in some detail. We take $\otimes$ to be the cartesian product, which is defined on relations $R:X\rarr X'$ and
$S:Y\rarr Y'$ as follows.
\[ \forall(x,y)\in X\times Y,(x',y')\in X'\times Y'.\; (x,y)R\otimes S(x',y')\iff xRx'\land ySy'\,. \]
It is not difficult to show that this is  indeed a functor. Note that, in the case that $R,S$ are \emph{functions}, $R\otimes S$ is the same as $R\times S$ in
\Set. Moreover, we take each $a_{A,B,C}$ to be the associativity function for products (in \Set), which is an iso in $\Set$ and hence also in \Rel.
Finally, we take $I$ to be the one-element set, and $l_A,r_A$ to be the projection functions: their relational converses are their inverses in \Rel.
The monoidal coherence diagrams commute simply because they commute in $\Set$.

\paragraph{Tensors and products}
As mentioned earlier, products are tensors with extra structure: natural diagonals and projections, corresponding to cloning and deleting operations. This fact is expressed more precisely as follows.

\begin{proposition}
Let $\CC$ be a monoidal category $(\CC,\otimes,I,a,l,r)$. The tensor $\otimes$ induces a product structure iff there exist natural diagonals and projections,
i.e.~natural transformations 
\[ \Delta_A:A\lrarr A\otimes A\,,\qquad p_{A,B}:A\otimes B\lrarr A\,,\qquad q_{A,B}:A\otimes B\lrarr B\,, \]
such that the following diagrams commute.
\[ \begin{diagram}
& & A & & \\
& \ldTo^{\id{A} } & \dTo_{\Delta_{A}} & \rdTo^{\id{A}} & \\
A & \lTo_{p_{A, A}} & A \otimes A & \rTo_{q_{A, A}} & A
\end{diagram} \qquad
\begin{diagram}[4em]
A \otimes B & \rTo^{\Delta_{A, B}} & (A \otimes B) \otimes (A \otimes B) \\
& \rdTo_{\id{A \otimes B}} & \dTo_{p_{A, B} \otimes q_{A, B}} \\
& & A \otimes B
\end{diagram}
\]
\end{proposition}

\paragraph{Symmetry} A \emph{symmetric monoidal category} is a monoidal category $(\CC,\otimes,I,a,l,r)$ with an additional natural isomorphism (\emph{symmetry}),
\[ \sigma_{A, B} : A \otimes B \iso B \otimes A \]
such that $\sigma_{B,A}=\sigma_{A,B}^{-1}$, and some additional coherence diagrams commute.

\subsection{Scalars}

Let $(\CC , \otimes , \II, l, a, l, r)$ be a  monoidal category .
We define a \emph{scalar} in $\CC$ to be a morphism $s : \II \rightarrow \II$, \ie an endomorphism of the tensor unit.

\begin{example}
In $\mathbf{FdVec}_{\mathbb{K}}$, linear maps $\mathbb{K} \to \mathbb{K}$ are uniquely
determined by the image of $1$, and hence correspond biuniquely to
elements of $\mathbb{K}\,$; composition corresponds to multiplication of 
scalars. In $\mathbf{Rel}$\index{category of relations}, there are just two scalars, corresponding
to the Boolean values $0$, $1$.
\end{example}

\noindent The (multiplicative) monoid of scalars is then just the endomorphism monoid $\CC (\II , \II )$.
The first key point is the elementary but beautiful observation by Kelly and Laplaza \cite{KL} that this monoid is always commutative.
\begin{lemma}
\label{scprop}
$\CC (\II , \II )$ is a commutative monoid
\end{lemma}
\begin{proof}
\[ \begin{diagram}
\II & \rTo^{r_{\II}^{-1}} & \II \otimes \II &  \req & \II \otimes \II & \rTo^{l_{\II}} & \II \\
\uTo^{s} & & \uTo^{s \otimes 1} & & \dTo_{1 \otimes t} & & \dTo_{t} \\
\II & \rTo^{r_{\II}^{-1}} & \II \otimes \II & \rTo^{s \otimes t} & \II \otimes \II & \rTo^{l_{\II}} & \II \\
\dTo^{t} & & \dTo^{1 \otimes t} & & \uTo_{s \otimes 1} & & \uTo_{s} \\
\II & \rTo_{l_{\II}^{-1}} & \II \otimes \II &  \rEq & \II \otimes \II & \rTo_{r_{\II}} & \II \\
\end{diagram}
\]
using the coherence equation $l_{\II} = r_{\II}$.
\end{proof}

The second point is that a good notion of \emph{scalar multiplication} exists at this level of generality.
That is, each scalar
$s:\II\to\II$ induces a natural transformation
\[ \begin{diagram}
s_A :A & \rTo^{\simeq} & \II \otimes \!A & \rTo^{s \otimes 1_A} & \II
\otimes\! A & \rTo^{\!\!\simeq\ } & A\,.\ \ \ \ \
\end{diagram}
\]
with the naturality square
\[ \begin{diagram}
A & \rTo^{s_A} & A \\
\dTo^{f} & & \dTo_{f} \\
B & \rTo_{s_B} & B \\
\end{diagram}
\]
We write $s \sdot f$ for $f \circ s_A=s_B\circ f$.
Note that
\[ \begin{array}{lcl}
1 \sdot f & = & f \label{sdotident} \\
s \sdot (t \sdot f) & = &  (s \circ t) \sdot f \label{sdotact}\\
(s \sdot g)\circ(t \sdot f) & =  & (s\circ t)\sdot(g\circ f) \label{sdotcomp}\\
(s \sdot f) \otimes (t \sdot g) & = &  (s \circ t) \sdot (f \otimes g) \label{sdotten}
\end{array}
\]
which exactly generalizes the multiplicative part of the usual properties of scalar multiplication.
Thus scalars act globally on the whole category.

\subsection{Compact Closed Categories}

A category {\bf C} is \em $*$-autonomous \em \cite{Barr} if
it is symmetric monoidal, and comes equipped with a full and
faithful functor 
\[
(\ )^*:{\bf C}^{op}\to{\bf C}
\]
such that a bijection
\[ 
{\bf C}(A\otimes B,C^*)\simeq {\bf C}(A,(B\otimes C)^*)
\]
exists which is natural in all variables. 
Hence a $*$-autonomous category is closed, with
\[
{A\multimap B:=(A\otimes B^*)^*}\,.
\]
These $*$-autonomous categories
provide a categorical semantics for the multiplicative
fragment of linear logic \cite{Seely}.

A \em compact closed category \em \cite{KL} is a
$*$-autonomous category with a self-dual tensor\index{compact closure},
i.e.~with natural isomorphisms
\[
u_{A,B}:(A\otimes B)^*\simeq A^*\otimes B^* \qquad u_{\II} : \II^* \simeq
\II\,.
\] 
It follows that 
\[
A\multimap B\simeq A^*\otimes B\,.
\]

\noindent A very different definition arises when one considers a symmetric monoidal
category as a one-object bicategory. In this context,
compact closure simply means that every object $A$, qua 1-cell of the
bicategory, has a specified adjoint \cite{KL}. 
\begin{definition}[Kelly-Laplaza]\label{def:compclos}\em 
A \em compact closed category \em is a symmetric monoidal
category in which to each object $A$ a \em dual object \em  
$A^*$, a \em unit \em 
\[
\eta_A:{\rm I}\to A^*\otimes A
\]
 and a \em  
counit \em 
\[
\epsilon_A:A\otimes A^*\to {\rm I}
\]
are assigned, in
such a way that the diagram
\begin{diagram}  
A&\rTo^{r^{-1}_A}&A\otimes{\rm
I}&\rTo^{1_A\otimes\eta_A}&A\otimes(A^*\otimes
A)\\ 
\dTo^{1_A}&&&&\dTo_{a_{A,A^*\!\!,A}}\\ 
A&\lTo_{l_A}&{\rm I}\otimes A&\lTo_{\epsilon_A\otimes
1_A}&(A\otimes A^*)\otimes A
\end{diagram}
and the dual one for $A^*$ both commute.  
\end{definition} 

\paragraph{Examples}The symmetric monoidal categories $({\bf Rel},\times)$ of sets, relations
and cartesian product\index{category of relations} and $({\bf FdVec}_\mathbb{K},\otimes)$ of finite-dimensional vector spaces over a field
$\mathbb{K}$, linear maps and tensor product
are both compact closed.  In $({\bf Rel},\times)$, we simply set $X^{*} = X$.  Taking a one-point set $\{ * \}$ as the unit for $\times$, and writing $R^{\cup}$ for the converse of a relation $R$:
\[
\eta_X=\epsilon_X^{\cup}=\{(*,(x,x))\mid x\in X\}\,.
\] 
For $({\bf FdVec}_\mathbb{K},\otimes)$, we take $V^{*}$ to be the dual space of linear functionals on $V$.
The unit and counit in $({\bf FdVec}_\mathbb{K},\otimes)$ are
\[
\eta_V:\mathbb{K}\to V^*\otimes V::1\mapsto\sum_{i=1}^{i=n}\bar{e}_i\otimes
e_i
\qquad
{\rm and}
\qquad
\epsilon_V:V\otimes
V^*\to\mathbb{K}::e_i\otimes\bar{e}_j\mapsto \bar{e}_{j}( e_{i})
\] 
where $n$ is the dimension of $V$, $\{e_i\}_{i=1}^{i=n}$ is a basis of
$V$ and $\bar{e}_i$ is the linear functional in $V^*$ determined by
$\bar{e}_{j}( e_{i}) = \delta_{ij}$.

\begin{definition}\label{def:name}\em
The \em name \em $\uu f\uuu$ and the \em coname \em $\dd f\ddd$ of a morphism $f:A\to B$ in a compact
closed category are
\begin{diagram} 
A^*\!\!\otimes\! A&\rTo^{1_{A^*}\!\!\otimes\! f}&A^*\!\otimes\! B&&&&&{\rm I}\\
\uTo^{\eta_A}&\ruTo_{\uu f\uuu}&&&&&\ruTo^{\dd f\ddd}&\uTo_{\epsilon_B} \\     
{\rm I}&&&&&A\!\otimes\! B^*&\rTo_{f\!\otimes\! 1_{B^*}}&B\!\otimes\! B^*&&
\end{diagram}
\end{definition}

For $R\in{\bf Rel}(X,Y)$ we have
\[
\uu R\uuu=\{(*,(x,y))\mid xRy,x\in X, y\in Y\}\quad{\rm and}\quad
\dd R\ddd=\{((x, y),*)\mid xRy,x\in X, y\in Y\}
\]

\noindent 
and for $f\in {\bf FdVec}_\mathbb{K}(V,W)$ with $(m_{ij})$ the matrix of
$f$ in bases
$\{e_i^V\}_{i=1}^{i=n}$ and $\{e_j^W\}_{j=1}^{j=m}$ of $V$ and $W$
respectively
\[
\uu f\uuu:\mathbb{K}\to V^*\otimes
W::1\mapsto\!\!\sum_{i,j=1}^{\!i,j=n,m\!}\!\!m_{ij}\cdot
\bar{e}_i^V\otimes e_j^W
\]
and
\[ 
\dd f\ddd:V\otimes
W^*\to\mathbb{K}::e_i^V\otimes\bar{e}_j^W\mapsto m_{ij}.
\]

\noindent Given $f:A\to B$ in any compact closed category ${\bf C}$  we can define $f^*:B^*\to A^*$ as
\begin{diagram}  
B^*&\rTo^{l_{B^*}^{-1}}&{\rm I}\otimes B^*&\rTo^{\eta_A\otimes 1_{B^*}}&A^*\otimes A\otimes
B^*\\ 
\dTo^{f^*}&&&&\dTo_{1_{A^*}\!\otimes f\otimes 1_{B^*}}\\   
A^*&\lTo_{r_{A^*}}&A^*\otimes {\rm I}&\lTo_{1_{A^*}\otimes \epsilon_B}&A^*\otimes
B\otimes B^*
\end{diagram}
This operation $(\ )^*$ is functorial and makes Definition \ref{def:compclos} coincide
with the one given at the beginning of this section. It then follows by 
\[
{\bf C}(A\otimes B^*,{\rm I}) \iso {\bf C}(A,B) \iso {\bf C}(I,A^*\otimes B)
\]
that every morphism
of type $\II\!\to\!   A^*\!\otimes B$ is the name of some morphism of type ${A\to B}$ and every
morphism of type ${A\otimes B^*\!\to{\rm I}}$ is the coname of some morphism of
type ${A\to B}$.  In the case of the unit and the counit we have
\[
\eta_A={\uu 1_A\uuu}\quad\quad{\rm and}
\quad\quad\epsilon_A={\dd 1_A\ddd}\,.  
\]
For $R\in{\bf Rel}(X,Y)$ the dual is the converse,
$R^*=R^{\cup}\in{\bf Rel}(Y,X)$, and for $f\in{\bf FdVec}_\mathbb{K}(V,W)$, the dual is
\[
f^*:W^*\to V^*::\,\phi \mapsto\,\phi\circ f\,.
\]

\subsection{Dagger Compact Categories}
In order to fully capture the salient structure of  $\FdHilb$, the category of finite-dimensional complex Hilbert spaces and linear maps, an important refinement of compact categories, to dagger- (or strongly-) compact categories, was introduced in \cite{AC2,AC3}. We shall not make any significant use of this refined definition in this paper, since our results hold at the more general level of compact categories.\footnote{We shall often use the abbreviated form ``compact categories'' instead of ``compact closed categories''.}
Nevertheless, we give the definition since we shall refer to this notion later.

We shall adopt the most  concise and elegant axiomatization of strongly or dagger compact closed categories, which takes the adjoint as primitive, following \cite{AC3}.
It is convenient to build the definition up in several stages, as in \cite{Selinger}.

\begin{definition}
A \emph{dagger category} is a category $\CC$ equipped with an identity-on-objects, contravariant, strictly involutive functor $f\mapsto f^\dagger$:
\[ 1^{\dagger} = 1, \qquad (g \circ f)^{\dagger} = f^{\dagger} \circ g^{\dagger}, \qquad f^{\dagger\dagger} = f \, . \]
We define an arrow $f : A \rarr B$ in a dagger category to be \emph{unitary} if it is an isomorphism such that $f^{-1} = f^{\dagger}$. An endomorphism $f : A \rarr A$ is \emph{self-adjoint} if $f = f^{\dagger}$.
\end{definition}

\begin{definition}
A \emph{dagger symmetric monoidal category} $(\CC , \otimes , \II, a, l, r, \sigma, {\dagger} )$
combines dagger and symmetric monoidal structure, with the requirement that the natural isomorphisms 
$a$, $l$, $r$, $\sigma$ are componentwise unitary, and moreover that $\dagger$ is a strict monoidal functor:
\[ (f \otimes g)^{\dagger} = f^{\dagger} \otimes g^{\dagger} \, .
\]
\end{definition}

Finally we come to the main definition.
\begin{definition}
A \emph{dagger compact  category} is a dagger symmetric monoidal category which is compact closed, and such that the following diagram commutes:
\[ \begin{diagram}
\II & \rTo^{\eta_{A}} & A^{*} \otimes A \\
& \rdTo_{\epsilon_{A}^{\dagger}} & \dTo_{\sigma_{A^{*}, A}} \\
& & A \otimes A^{*}
\end{diagram}
\]
\end{definition}
\noindent This implies that  the counit is \emph{definable} from the unit and the adjoint: 
\[ \epsilon_{A} =  \eta_{A}^{\dagger} \circ \sigma_{A, A^{*}} \]
and similarly the unit can be defined from the counit and the adjoint.
Furthermore, it is in fact possible to replace the two commuting diagrams required in the definition of compact closure by one. We refer to \cite{AC3} for the details.

\subsection{Trace}

An essential mathematical instrument in quantum mechanics is the \emph{trace} of a linear map. In quantum information, extensive use is made of the more general notion of \emph{partial trace}, which is used to trace out a subsystem of a compound system. 

A general categorical axiomatization of the notion of partial trace has been given by Joyal, Street and Verity \cite{JSV}. A trace in a symmetric monoidal category $\CC$ is a family of functions 
\[ \Tr_{A,B}^U : \CC (A \otimes U, B \otimes U) \longrightarrow \CC (A, B) \]
for objects $A$, $B$, $U$ of $\CC$, satisfying a number of axioms, for which we refer to \cite{JSV}.
This specializes to yield the total trace for endomorphisms by taking $A = B = \II$. In this case, $\Tr(f) = \Tr_{\II,\II} ^{U}(f): \II \rarr \II$ is a scalar. Expected properties such as the invariance of the trace under cyclic permutations
\[ \Tr(g \circ f) = \Tr(f \circ g) \]
follow from the general axioms.

Any compact closed category carries a canonical (in fact, a unique) trace.  For an endomorphism $f : A \rarr A$, the total trace is defined by
\[ \Tr(f) = \epsilon_{A} \circ (f \otimes 1_{A^{*}}) \circ \sigma_{A^{*},A} \circ \eta_{A} \, . \]
This definition gives rise to the standard notion of trace  in $\FdHilb$.

\subsection{Graphical Representation}

Complex algebraic expressions for morphisms in symmetric monoidal categories can rapidly become hard to read.  Graphical representations exploit two-dimensionality, with the vertical dimension  corresponding to composition and the horizontal to the monoidal tensor,
and provide more intuitive presentations of morphisms.  We depict  objects by wires, morphisms by boxes with input and output wires, composition by connecting outputs to inputs, and the monoidal tensor by locating boxes side-by-side.
\begin{center}
\centering{\epsfig{figure=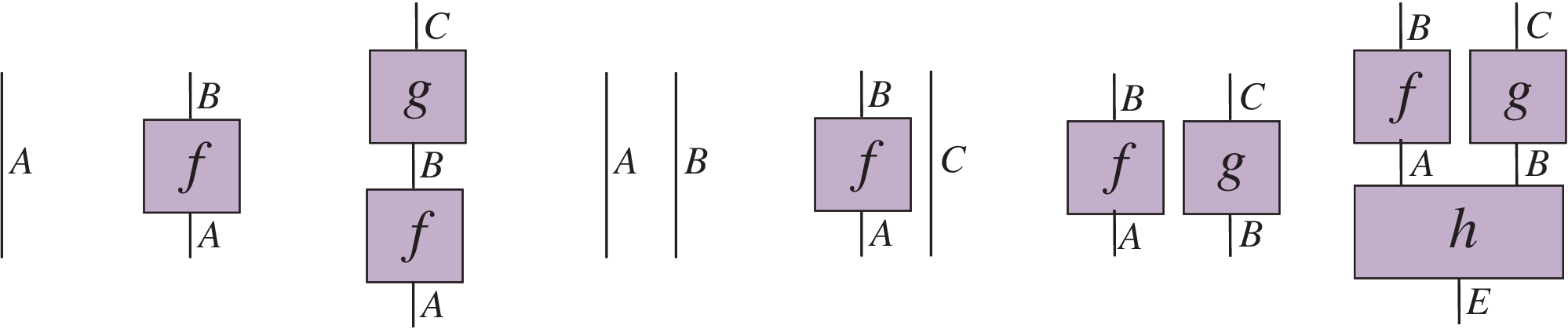,width=280pt}}      
\end{center}

\noindent Algebraically, these correspond to:
\[ 1_A:A\to A, \quad f:A\to B, \quad g\circ f, \quad 1_A\otimes 1_B, \quad f\otimes 1_C, \quad f\otimes g, \quad (f\otimes g)\circ h \]
respectively.
(The convention in these diagrams is that the `upward' vertical direction represents progress of time.)

\paragraph{Kets, Bras and Scalars:}
A special role is played by boxes with either no input or no output,  \ie arrows of the form $I \lrarr A$ or $A \lrarr I$ respectively, where $I$ is the unit of the tensor. In the setting of $\FdHilb$ and Quantum Mechanics, they correspond to \emph{states} and \emph{costates} respectively (cf.~Dirac's kets and bras  \cite{Dirac}), which we depict by triangles. \emph{Scalars} then arise naturally by composing these elements (cf.~inner-product or Dirac's bra-ket):
\begin{center}
\centering{\epsfig{figure=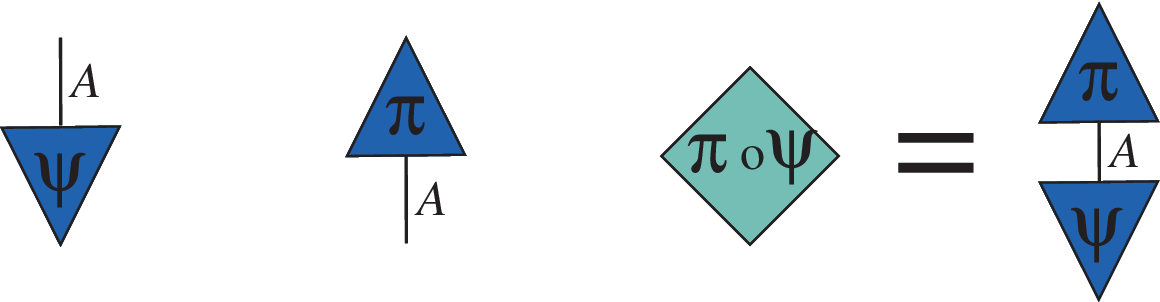,width=200pt}} 
\end{center}
Formally, scalars are arrows of the form $I \lrarr I$. In the physical context, they provide 
numbers (``probability amplitudes'' etc.). For example, in $\FdHilb$, the tensor unit is $\Complex$, the complex numbers, and a linear map $s : \Complex \lrarr \Complex$ is determined by a single number, $s(1)$. In $\Rel$, the scalars are the boolean semiring $\{ 0, 1 \}$.

This graphical notation can be seen as a substantial two-dimensional generalization of \emph{Dirac notation} \cite{Dirac}:
\[ \langle \phi \mid \qquad \qquad \mid \psi \rangle \qquad \qquad \langle \phi \mid \psi \rangle \]
Note how the geometry of the plane absorbs functoriality and naturality conditions, e.g.:
\begin{center}
\psset{unit=1in,cornersize=absolute,dimen=middle}%
\begin{pspicture}(0,-1)(4,1)%
% dpic version 08.Jul.05 for PSTricks 0.93a
\psframe[fillstyle=solid,fillcolor=yellow,linecolor=blue](0,0.2)(0.4,0.6)
\rput(0.2,0.4){$f$}
\psline(0.2,0.6)(0.2,1)
\psline(0.2,0.2)(0.2,-1)
\psframe[fillstyle=solid,fillcolor=yellow,linecolor=blue](0.8,-0.6)(1.2,-0.2)
\rput(1,-0.4){$g$}
\psline(1,-0.2)(1,1)
\psline(1,-0.6)(1,-1)
\rput(2,0){$=$}
\psframe[fillstyle=solid,fillcolor=yellow,linecolor=blue](2.8,-0.6)(3.2,-0.2)
\rput(3,-0.4){$f$}
\psline(3,-0.2)(3,1)
\psline(3,-0.6)(3,-1)
\psframe[fillstyle=solid,fillcolor=yellow,linecolor=blue](3.6,0.2)(4,0.6)
\rput(3.8,0.4){$g$}
\psline(3.8,0.6)(3.8,1)
\psline(3.8,0.2)(3.8,-1)
\end{pspicture}%
\end{center}

\vspace{-.2in}
\[ \;\; (f \otimes 1) \circ (1 \otimes g) \qquad   =  \qquad f \otimes g \qquad =   \qquad (1 \otimes g) \circ (f \otimes 1)  \]

%\vspace{.2in}
\paragraph{Cups and Caps}
We introduce a special diagrammatic notation for the unit and counit.
\vspace{-.3in}
\begin{center}
\psset{unit=1in,cornersize=absolute,dimen=middle}%
\begin{pspicture}(0,-1.263158)(3.157895,0.736842)%
% dpic version 08.Jul.05 for PSTricks 0.93a
\psset{linewidth=1pt}%
\pscustom[fillcolor=lightgray,fillstyle=solid,linecolor=blue]{%
\psline(0,-0.315789)(1.052632,-0.315789)
(0.526316,0.210526)
(0,-0.315789)
}%
\psline[arrowsize=0.05in 0,arrowlength=2,arrowinset=0]{->}(0.263158,-0.666667)(0.263158,-0.22807)
(0.789474,-0.22807)
(0.789474,-0.666667)
\rput(0.789474,-0.736842){$A^*$}
\rput(0.263158,-0.736842){$A$}
\psset{linewidth=0.8pt}%
\psset{linewidth=1pt}%
\pscustom[fillcolor=yellow,fillstyle=solid,linecolor=blue]{%
\psline(2.105263,-0.210526)(3.157895,-0.210526)
(2.631579,-0.736842)
(2.105263,-0.210526)
}%
\psline[arrowsize=0.05in 0,arrowlength=2,arrowinset=0]{->}(2.368421,0.140351)(2.368421,-0.298246)
(2.894737,-0.298246)
(2.894737,0.140351)
\rput(2.368421,0.210526){$A^*$}
\rput(2.894737,0.210526){$A$}
\psset{linewidth=0.8pt}%
\end{pspicture}%
\end{center}
\vspace{-.4in}
\[  \epsilon_{A} : A \otimes A^{*} \longrightarrow \II \qquad \qquad  \qquad  \qquad \eta_{A} : \II \longrightarrow A^{*} \otimes A . \]
The lines indicate the \emph{information flow} accomplished by these operations.

\paragraph{Compact Closure}
The basic algebraic laws for units and counits become diagrammatically evident in terms of the information-flow lines:
%\vspace{-.2in}
\begin{center}
\psset{unit=1in,cornersize=absolute}%
\begin{pspicture}(0,-1.329545)(4.5,0.306818)
% dpic version 28.Jul.03 for PSTricks 0.93a
\psset{linewidth=1pt}%
\pscustom[fillcolor=lightgray,fillstyle=solid,linecolor=blue]{%
\psline(0,-0.306818)(0.613636,-0.306818)
(0.306818,0)
(0,-0.306818)
}%
\pscustom[fillcolor=yellow,fillstyle=solid,linecolor=blue]{%
\psline(0.409091,-0.715909)(1.022727,-0.715909)
(0.715909,-1.022727)
(0.409091,-0.715909)
}%
\psset{linewidth=2pt}%
\psline[arrowsize=0.1in 0,arrowlength=1.25,arrowinset=0]{->}(0.102273,-1.022727)(0.102273,-0.255682)
(0.511364,-0.255682)
(0.511364,-0.767045)
(0.920455,-0.767045)
(0.920455,0)
\rput(1.431818,-0.306818){{\Huge $=$}}
\psset{linewidth=2pt}%
\psline[arrowsize=0.1in 0,arrowlength=1.25,arrowinset=0]{->}(1.840909,-1.022727)(1.840909,0)
\psset{linewidth=1pt}%
\pscustom[fillcolor=lightgray,fillstyle=solid,linecolor=blue]{%
\psline(3.068182,-0.306818)(3.681818,-0.306818)
(3.375,0)
(3.068182,-0.306818)
}%
\pscustom[fillcolor=yellow,fillstyle=solid,linecolor=blue]{%
\psline(2.659091,-0.715909)(3.272727,-0.715909)
(2.965909,-1.022727)
(2.659091,-0.715909)
}%
\psset{linewidth=2pt}%
\psline[arrowsize=0.1in 0,arrowlength=1.25,arrowinset=0]{->}(2.761364,0)(2.761364,-0.767045)
(3.170455,-0.767045)
(3.170455,-0.255682)
(3.579545,-0.255682)
(3.579545,-1.022727)
\rput(4.090909,-0.306818){{\Huge $=$}}
\psset{linewidth=2pt}%
\psline[arrowsize=0.1in 0,arrowlength=1.25,arrowinset=0]{->}(4.5,0)(4.5,-1.022727)
\end{pspicture}%
\end{center}
\vspace{-.2in}
\[ \qquad (\epsilon_A \otimes 1_A ) \circ (1_A \otimes \eta_A ) = 1_A  \quad \qquad  \qquad
(1_{A^*} \otimes \epsilon_A ) \circ (\eta_A \otimes 1_{A^*} ) = 1_{A^*}
\]

\paragraph{Names and Conames in the Graphical Calculus}
The units and counits are powerful; they allow us to define a \emph{closed structure} on the category.
In particular, we can form the \emph{name} $\uu f \uuu$ of any arrow $f : A \rightarrow B$, as a special case  of $\lambda$-abstraction, and dually the \emph{coname} $\dd f \ddd$:
\begin{center}
\psset{unit=1in,cornersize=absolute}%
\begin{pspicture}(0,-1.555556)(2.962963,0.444444)
% dpic version 28.Jul.03 for PSTricks 0.93a
\psset{linewidth=1pt}%
\pscustom[fillcolor=lightgray,fillstyle=solid,linecolor=blue]{%
\psline(0,-0.444444)(0.888889,-0.444444)
(0.444444,0)
(0,-0.444444)
}%
\psframe[fillstyle=solid,fillcolor=lightgray,linecolor=blue](-0.0041,-0.818915)(0.300396,-0.514419)
\rput(0.148148,-0.666667){{ $f$}}
\psset{linewidth=2pt}%
\psline(0.148148,-1.111111)(0.148148,-0.814815)
\psline[arrowsize=0.1in 0,arrowlength=1.25,arrowinset=0]{->}(0.148148,-0.518519)(0.148148,-0.37037)
(0.740741,-0.37037)
(0.740741,-1.111111)
\psset{linewidth=1pt}%
\pscustom[fillcolor=yellow,fillstyle=solid,linecolor=blue]{%
\psline(2.074074,-0.666667)(2.962963,-0.666667)
(2.518519,-1.111111)
(2.074074,-0.666667)
}%
\psframe[fillstyle=solid,fillcolor=yellow,linecolor=blue](2.662567,-0.596692)(2.967063,-0.292196)
\rput(2.814815,-0.444444){{ $f$}}
\psset{linewidth=2pt}%
\psline(2.814815,0)(2.814815,-0.296296)
\psline[arrowsize=0.1in 0,arrowlength=1.25,arrowinset=0]{->}(2.814815,-0.592593)(2.814815,-0.740741)
(2.222222,-0.740741)
(2.222222,0)
\end{pspicture}%
\end{center}
\vsn\vsn
\[ \dd f \ddd : A \otimes B^* \rightarrow \II \quad \quad \quad \quad \quad \quad \;\; \uu f \uuu : \II \rightarrow A^* \otimes B \]
This is the general form of Map-State duality:
\[ \CC (A \otimes B^\ast , \II ) \iso \CC (A, B) \iso \CC (\II ,
A^\ast \otimes B).
\]

\subsection{Formalizing Quantum Information Flow}

In this section, we give a brief glimpse of  categorical quantum mechanics. While not needed for the results to follow, it provides the motivating context for them. For further details, see e.g. \cite{AC4}.

\subsubsection{Quantum Entanglement}
We consider for illustration two standard examples of two-qubit entangled states,
the Bell state:
\begin{center}
\psset{unit=1in,cornersize=absolute}%
\begin{pspicture}(0,-0.1)(2.4,0.1)
% dpic version 04.Aug.03 for PSTricks 0.93a
\psset{linewidth=2pt}%
\pscircle[fillstyle=solid,fillcolor=black](0.05,0){0.1}
\psline(0.1,0)(2,0)
\uput{0.5ex}[u](1,0.05){\mbox{{$|00\rangle + |11\rangle$}}}
\pscircle[fillstyle=solid,fillcolor=black](2,0){0.1}
\end{pspicture}%
\end{center}
\vspace{.2in}

\noindent and the EPR state:
\begin{center}
\psset{unit=1in,cornersize=absolute}%
\begin{pspicture}(0,-0.1)(2.4,0.1)
% dpic version 04.Aug.03 for PSTricks 0.93a
\psset{linewidth=2pt}%
\pscircle[fillstyle=solid,fillcolor=black](0.05,0){0.1}
\psline(0.1,0)(2,0)
\uput{0.5ex}[u](1,0.05){\mbox{{$|01\rangle + |10\rangle$}}}
\pscircle[fillstyle=solid,fillcolor=black](2,0){0.1}
\end{pspicture}%
\end{center}
\vspace{.1in}

In quantum mechanics, compound  systems are represented by  the \emph{tensor product} of Hilbert spaces: 
$\HH_1\otimes \HH_2$.
A typical element of the tensor product has the form: 
\[\sum_{i} \lambda_i \cdot \phi_i \otimes \psi_i \]
where $\phi_{i}$, $\psi_{i}$ range over basis vectors, and the coefficients $\lambda_{i}$ are complex numbers.
\emph{Superposition} encodes \emph{correlation}: in the Bell state, the off-diagonal elements have zero coefficients.
This gives rise to
Einstein's ``spooky action at a distance''.
Even if the particles are spatially separated, measuring one has an
effect on the state of the other.
In the Bell state, for example, when we measure one of the two qubits we may get either 0 or 1, but once this result has been obtained, it is certain that the result of measuring the other qubit will be the same.

This leads to
Bell's famous theorem \cite{Bell}: QM is \emph{essentially non-local}, in the sense that the correlations it predicts exceed those of any ``local realistic theory''.

\paragraph{From `paradox' to `feature': Teleportation}
%\vspace{.2in}
\begin{center}
\psset{unit=1in,cornersize=absolute}%
\begin{pspicture}(0,-1.062458)(3.88525,1.437542)
% dpic version 28.Jul.03 for PSTricks 0.93a
\psframe[fillstyle=solid,fillcolor=yellow,linecolor=blue](-0.002688,-0.196951)(1.168263,0.196951)
\rput(0.582787,0){{ $M_{\mathrm{Bell}}$}}
\psframe[fillstyle=solid,fillcolor=green,linecolor=blue](3.494037,0.580099)(3.887938,0.974)
\rput(3.690987,0.77705){{ $U_x$}}
\newgray{fillval}{0.3}
\pscircle[fillstyle=solid,fillcolor=fillval](0.291394,-0.971312){0.084623}
\pscircle[fillstyle=solid,fillcolor=fillval](0.874181,-0.971312){0.084623}
\pscircle[fillstyle=solid,fillcolor=fillval](3.690987,-0.971312){0.084623}
\psset{linewidth=1pt}%
\psline[arrowsize=0.05in 0,arrowlength=2,arrowinset=0]{->}(0.291394,-0.893607)(0.291394,-0.194262)
\psline[arrowsize=0.05in 0,arrowlength=2,arrowinset=0]{->}(0.874181,-0.893607)(0.874181,-0.194262)
\psline[arrowsize=0.05in 0,arrowlength=2,arrowinset=0]{->}(3.690987,-0.893607)(3.690987,0.582787)
\psline[arrowsize=0.05in 0,arrowlength=2,arrowinset=0]{->}(3.690987,0.971312)(3.690987,1.359837)
\psset{linewidth=2pt}%
\psline(0.951886,-0.971312)(3.613282,-0.971312)
\uput{0.5ex}[u](2.282584,-0.971312){{ $| 00\rangle + | 11\rangle$}}
\psset{linewidth=1pt}%
\psline[linestyle=dashed,arrowsize=0.05in 0,arrowlength=2,arrowinset=0]{->}(0.582787,0.194262)(0.582787,0.77705)
(3.496725,0.77705)
\uput{0.5ex}[u](2.039756,0.77705){{ $x \in \mathbb{B}^2$}}
\uput{0.5ex}[d](0.291394,-1.049017){{ $| \phi \rangle$}}
\rput(3.690987,1.437542){{ $| \phi \rangle$}}
\end{pspicture}% 
\end{center}
\[ \qquad \quad \mbox{Alice} \qquad \qquad \qquad \qquad \qquad \qquad  \qquad \qquad  \qquad \qquad \mbox{Bob} \]

\vsa
\noindent In the teleportation protocol \cite{BBC}, Alice sends an unknown qubit $\phi$ to Bob, using a shared Bell pair as a ``quantum channel''. By performing a measurement in the Bell basis on $\phi$ and her half of the entangled pair, a collapse is induced on Bob's qubit. Once the result $x$ of Alice's measurement is transmitted by classical communication to Bob (there are four possible measurement outcomes, hence this requires two classical bits), Bob can perform a corresponding unitary correction $U_{x}$ on his qubit, after which it will be in the state $\phi$.

\subsubsection{Categorical Quantum Mechanics and Diagrammatics}
We now outline the categorical approach to quantum mechanics developed in \cite{AC2,AC3}.
The \emph{same} graphical calculus and underlying algebraic structure which we have seen in the previous section  has been  applied to quantum information and computation, yielding an incisive analysis of \emph{quantum information flow}, and powerful and illuminating methods for reasoning about quantum informatic processes and protocols \cite{AC2}.

\paragraph{Bell States and Costates:}
The cups and caps we have already seen in the guise of deficit and cancellation operations, now take on the r\^ole of \emph{Bell states and costates} (or preparation and test of Bell states), the fundamental building blocks of quantum entanglement. (Mathematically, they arise as the transpose and co-transpose of the identity, which exist in any finite-dimensional Hilbert space by ``map-state duality''). 
\begin{center}
\centering{\epsfig{figure=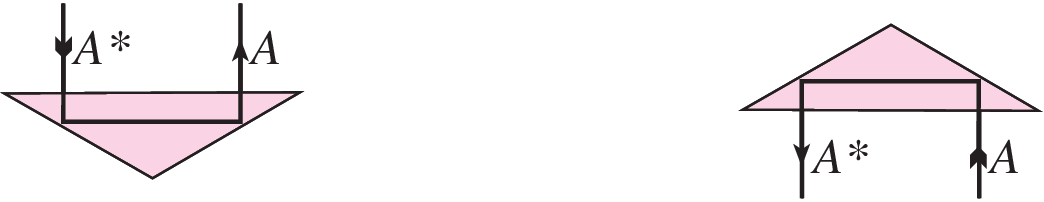,width=200pt}} 
\end{center}

\noindent The formation of \emph{names} and \emph{conames} of arrows (\ie map-state and map-costate duality) is conveniently depicted thus:

\begin{center}
\centering{\fbox{\epsfig{figure=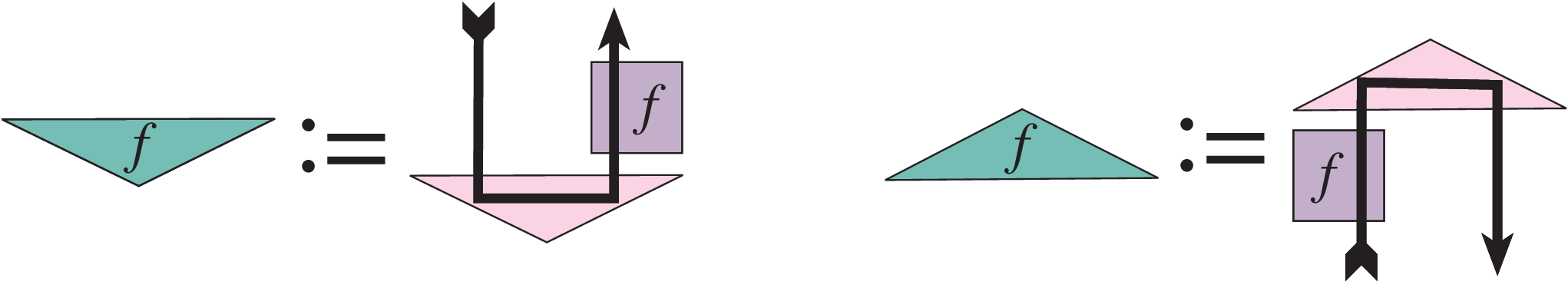,width=285pt}\hfill{\bf (2)}}}     
\end{center}

\noindent The key lemma in exposing the quantum information flow in (bipartite) entangled quantum systems can be formulated diagrammatically as follows:

\begin{minipage}[b]{1\linewidth}
\centering{\epsfig{figure=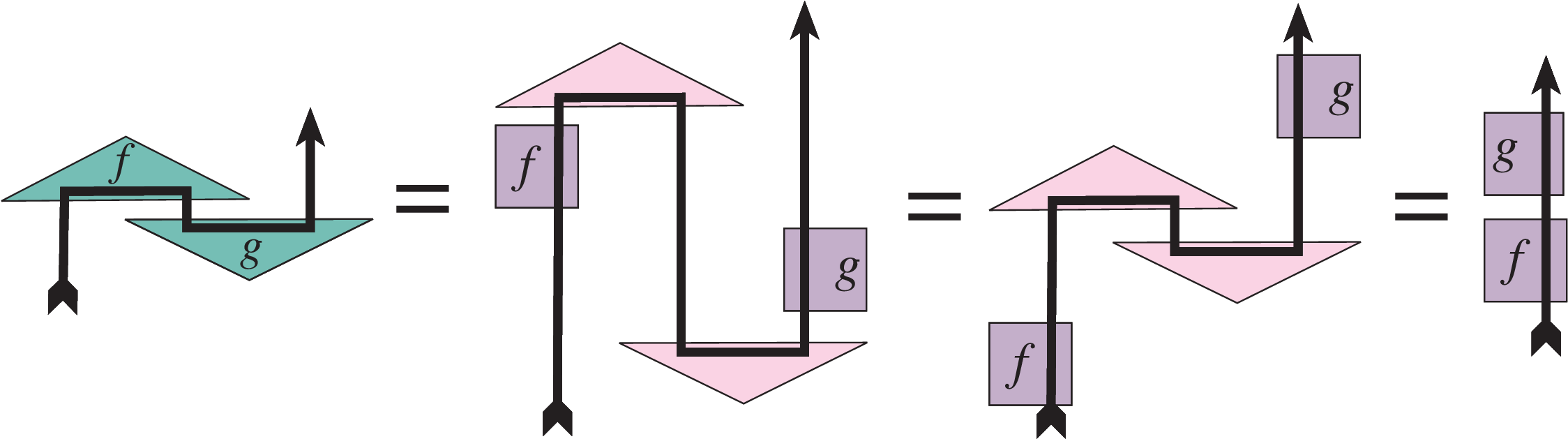,width=300pt}}     
\end{minipage}
Note in particular the interesting phenomenon of ``apparent reversal of the causal order'' .
While on the left, physically, we first prepare the state labeled $g$ and then apply the costate labeled $f$, the global effect is {\em as if} we first applied $f$ itself first, and only then $g$. 

\paragraph{Derivation of quantum teleportation.} This is the most basic application of compositionality in action.  We can read off the basic quantum mechanical potential for teleportation immediately from the geometry of Bell states and costates:

\bigskip\noindent
\begin{minipage}[b]{1\linewidth}
\centering{$\!\!$\epsfig{figure=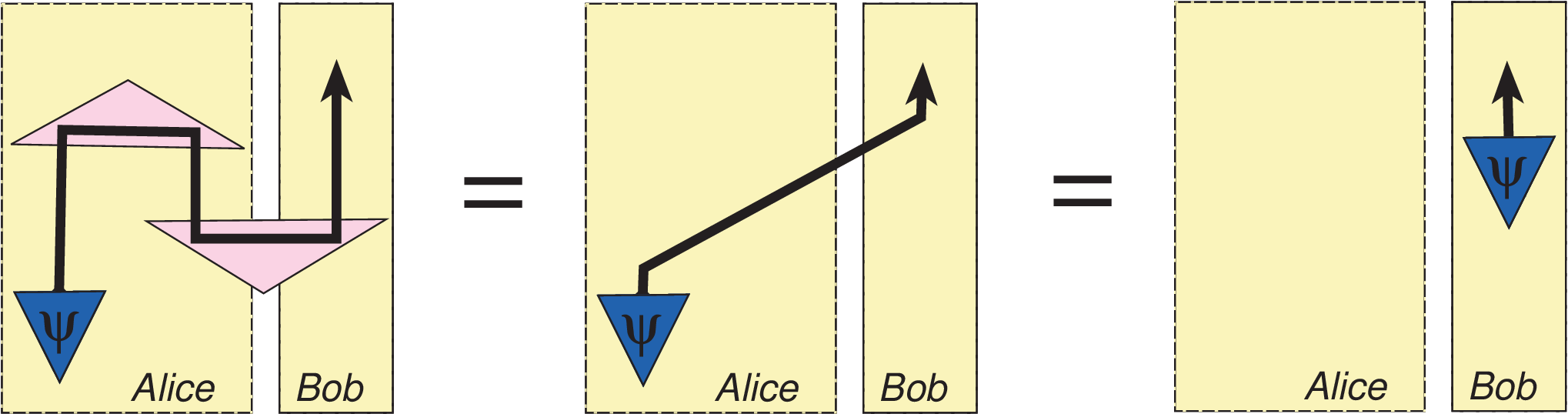,width=300pt}}     
\end{minipage}

\noindent
The Bell state forming the shared channel between Alice and Bob appears as the downwards triangle in the diagram; the Bell costate forming one of the possible measurement branches is the upwards triangle. The information flow of the input qubit from Alice to Bob is then immediately evident from the diagrammatics.

This is not quite the whole story, because of  the non-deterministic nature of measurements.
But in fact, allowing for this shows the underlying \emph{design principle} for the teleporation protocol. Namely, we find a measurement basis such that each possible branch $i$ through the measurement is labelled, under map-state duality, with a unitary map $f_{i}$. The corresponding correction is then just the inverse map $f_{i}^{-1}$. Using our lemma,
the full description of teleportation becomes:

\begin{center}
\centering{\epsfig{figure=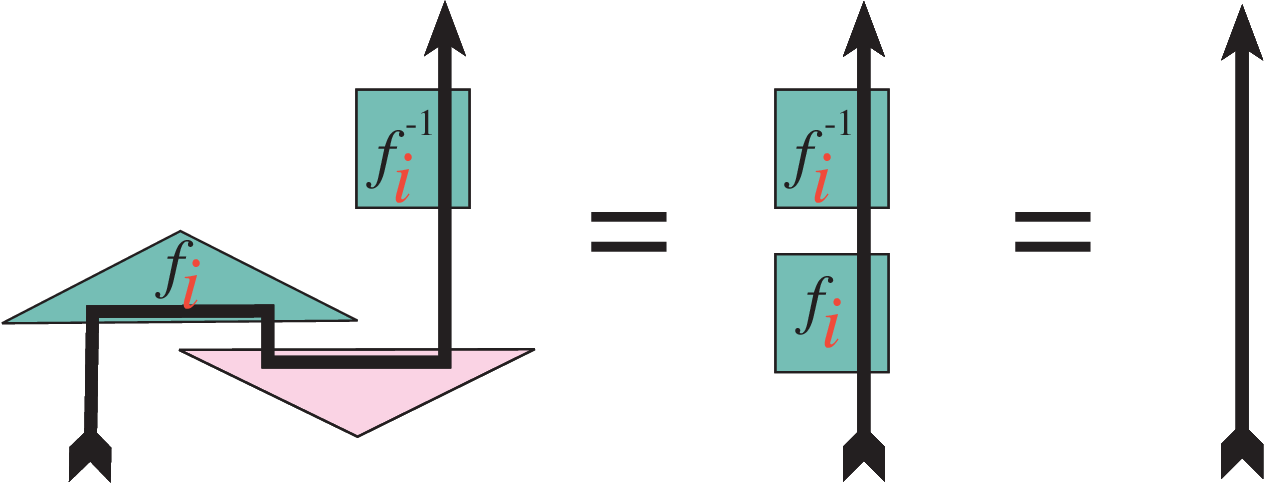,width=220pt}}
\end{center}

\section{No-Cloning}

Note that the proof of Joyal's lemma given in Section~\ref{JLsec} makes full use of both diagonals and projections, \ie of both cloning and deleting. Our aim  is to examine cloning and deleting as separate principles, and to see how far each in isolation is compatible with the strong form of duality which, as we have seen, plays a basic structural r\^ole in the categorical axiomatization of quantum mechanics, and applies very directly to the analysis of entanglement.

\subsection{Axiomatizing Cloning}

Our first task is to axiomatize cloning as a \emph{uniform operation} in the setting of a symmetric monoidal category.

As a preliminary, we recall the notions of monoidal functor and monoidal natural transformation.
Let $\CC$ and $\DD$ be monoidal categories.
A (strong) monoidal functor $(F, e, m) : \CC \lrarr \DD$ comprises:
\begin{itemize}
\item A functor $F : \CC \lrarr \DD$
\item An isomorphism $e : I \cong FI$
\item A natural isomorphism $m_{A, B} : FA \otimes FB \lrarr F(A \otimes B)$
\end{itemize}
subject to various coherence conditions.

Let $(F, e, m), (G, e', m') : \CC \lrarr \DD$ be monoidal functors.
A monoidal natural transformation between them is a natural transformation
$t : F \natarrow G$ such that
\[
\begin{diagram}
I & \rTo^{e} & FI \\
& \rdTo_{e'} & \dTo_{t_{I}} \\
& & GI
\end{diagram}
\qquad \qquad
\begin{diagram}
FA \otimes FB & \rTo^{m_{A, B}} & F(A \otimes B) \\
\dTo^{t_{A} \otimes t_{B}} & & \dTo_{t_{A \otimes B}} \\
GA \otimes GB & \rTo_{m'_{A, B}} & G(A \otimes B)
\end{diagram}
\]

\noindent We say that a monoidal category \emph{has uniform cloning} it is has a diagonal, \ie a monoidal natural transformation
\[ \Delta_{A} : A \lrarr A \otimes A \]
which is moreover \emph{coassociative and cocommutative}:
\[
\begin{diagram}
A & \rTo^{\Delta} & A \otimes A & \rTo^{1 \otimes \Delta} & A \otimes (A \otimes A) \\
\deq & & & & \dTo_{a_{A,A,A}} \\
A & \rTo_{\Delta} &  A \otimes A & \rTo_{\Delta \otimes 1} & (A \otimes A) \otimes A)
\end{diagram}
\qquad \qquad
\begin{diagram}
A & \rTo^{\Delta} & A \otimes A \\
& \rdTo_{\Delta} & \dTo_{\sigma_{A,A}} \\
& & A \otimes A
\end{diagram}
\]
Note that in the case when the monoidal structure is induced by a product, the standard diagonal
\[ \Delta_{A} : A \rTo^{\ang{1_{A}, 1_{A}}} A \times A \]
automatically satisfies all these properties.

To simplify the presentation, we shall henceforth make the assumption that the monoidal categories we consider are \emph{strictly associative}. This is a standard manouevre, and by the coherence theorem for monoidal categories \cite{Mac} is harmless.

Note that the functor $A \mapsto A \otimes A$ which is the codomain of the diagonal has as its monoidal structure maps
\[ 
\begin{diagram}
m_{A,B}   = A \otimes B \otimes A \otimes B & \rTo^{1 \otimes \sigma \otimes 1} & A \otimes A \otimes B \otimes B, & \quad &
e = \II & \rTo^{l_{\II}^{-1}} & \II \otimes \II \, .
\end{diagram}
\]
Of course the identity functor, which is the domain of the diagonal, has identity morphisms as its structure maps.

\subsection{Compact categories with cloning (almost) collapse}

\begin{theorem}
\label{thethe}
Let $\CC$ be a compact category with cloning.
Then every endomorphism is a scalar multiple of the identity. More precisely, for $f : A \rarr A$, $f = \Tr(f) \sdot \id{A}$. This means that  for every object $A$ of $\CC$, $\CC(A, A)$ is a retract of $\CC(\II, \II)$:
\[ \alpha : \CC(A, A) \lhd \CC(\II, \II) : \beta, \qquad  \alpha(f) = \Tr(f), \quad \beta(s) =  s \sdot \id{A} \, . \]
\end{theorem}
In a category enriched over vector spaces, this means that each endomorphism algebra is one-dimensional.
In the cartesian case, there is a unique scalar, and we recover the reflexive part of the posetal collapse of Joyal's lemma. But in general, the collapse given by our result is of a different nature to that of Joyal's lemma, as we shall see later.

Note that our collapse result only refers to endomorphisms. In the dagger-compact case, every morphism $f : A \rarr B$ has an associated endomorphism 
\[ \sigma \circ (f \otimes f^{\dagger}) : A \otimes B \rarr A \otimes B \, . \]
Moreover the passage to this associated endomorphism can be seen as a kind of ``projective quotient'' of the original category \cite{deLL}. Thus in this case, the collapse given by our theorem can be read as saying that the projective quotient of the category is trivial.

\subsection{Proving the Cloning Collapse Theorem}

We shall make some use of the graphical calculus in our proofs. We shall use slightly different conventions from those adopted in the previous section:
\begin{itemize}
\item Firstly, the diagrams to follow are to be read downwards rather than upwards.
\item Secondly, we shall depict the units and counits of a compact category simply as ``cups'' and ``caps'', without any enclosing triangles.
\end{itemize}
To illustrate these points, the units and counits will be depicted thus:

\vsa
\begin{center}
\psset{unit=1in,cornersize=absolute,dimen=middle}%
\begin{pspicture}(0,-0.478027)(2.25,0.126465)%
% dpic version 28.Jun.06 for PSTricks 0.93a
\psset{linewidth=1pt}%
\newgray{fillval}{0.3}
\pscircle[fillstyle=solid,fillcolor=fillval](0.017578,-0.084473){0.017578}
\pscircle[fillstyle=solid,fillcolor=fillval](0.404297,-0.084473){0.017578}
\psarcn(0.210938,-0.066895){0.193359}{-180}{-360}
\uput{0.501875ex}[d](0.017578,-0.102051){$A^*$}
\uput{0.501875ex}[d](0.404297,-0.102051){$A$}
\pscircle[fillstyle=solid,fillcolor=fillval](1.845703,0.084473){0.017578}
\pscircle[fillstyle=solid,fillcolor=fillval](2.232422,0.084473){0.017578}
\psarc(2.039063,0.066895){0.193359}{180}{360}
\uput{0.501875ex}[u](1.845703,0.102051){$A$}
\uput{0.501875ex}[u](2.232422,0.102051){$A^*$}
\rput(0.210937,-0.478027){$\eta_{A} : \II \longrightarrow A^{*} \otimes A$}
\rput(2.039063,-0.478027){$ \epsilon_{A} : A \otimes A^{*} \longrightarrow \II$}
\end{pspicture}%
\end{center}

\noindent while the identities for the units and counits in compact categories will appear thus:
\begin{center}
\psset{unit=1in,cornersize=absolute,dimen=middle}%
\begin{pspicture}(0,-0.733696)(2.184783,0.016304)%
% dpic version 08.Jul.05 for PSTricks 0.93a
\psset{linewidth=1pt}%
\newgray{fillval}{0.3}
\pscircle[fillstyle=solid,fillcolor=fillval](0.016304,0){0.016304}
\pscircle[fillstyle=solid,fillcolor=fillval](0.016304,-0.358696){0.016304}
\pscircle[fillstyle=solid,fillcolor=fillval](0.211957,-0.358696){0.016304}
\pscircle[fillstyle=solid,fillcolor=fillval](0.407609,-0.358696){0.016304}
\pscircle[fillstyle=solid,fillcolor=fillval](0.407609,-0.717391){0.016304}
\pscircle[fillstyle=solid,fillcolor=fillval](1.092391,-0.717391){0.016304}
\pscircle[fillstyle=solid,fillcolor=fillval](1.777174,-0.717391){0.016304}
\pscircle[fillstyle=solid,fillcolor=fillval](1.777174,-0.358696){0.016304}
\pscircle[fillstyle=solid,fillcolor=fillval](1.972826,-0.358696){0.016304}
\pscircle[fillstyle=solid,fillcolor=fillval](2.168478,-0.358696){0.016304}
\pscircle[fillstyle=solid,fillcolor=fillval](1.092391,-0){0.016304}
\pscircle[fillstyle=solid,fillcolor=fillval](2.168478,-0){0.016304}
\psarcn(0.309783,-0.342391){0.097826}{180}{0}
\psarcn(1.875,-0.342391){0.097826}{179.999998}{0.000002}
\psarc(0.11413,-0.375){0.097826}{-179.999999}{-0.000001}
\psarc(2.070652,-0.375){0.097826}{-179.999998}{-0.000002}
\psline(0.016304,-0.016304)(0.016304,-0.342391)
\psline(0.407609,-0.375)(0.407609,-0.701087)
\psline(1.092391,-0.016304)(1.092391,-0.701087)
\psline(1.777174,-0.375)(1.777174,-0.701087)
\psline(2.168478,-0.016304)(2.168478,-0.342391)
\rput(0.75,-0.358696){$=$}
\rput(1.402174,-0.358696){$=$}
\end{pspicture}%
\end{center}
The small nodes appearing in these diagrams indicate how the figures are built by composition from basic figures such as cups, caps and identities.
\paragraph{First step}
We shall begin by showing that
\begin{center}
``parallel caps = nested caps''
\end{center}

\noindent Diagrammatically:
\begin{center}
\psset{unit=1in,cornersize=absolute,dimen=middle}%
\begin{pspicture}(0,-0.036779)(2.5,0.413043)%
% dpic version 08.Jul.05 for PSTricks 0.93a
\psset{linewidth=1pt}%
\newgray{fillval}{0.3}
\pscircle[fillstyle=solid,fillcolor=fillval](0.021739,0){0.021739}
\pscircle[fillstyle=solid,fillcolor=fillval](0.282609,0){0.021739}
\pscircle[fillstyle=solid,fillcolor=fillval](0.543478,0){0.021739}
\pscircle[fillstyle=solid,fillcolor=fillval](0.804348,0){0.021739}
\pscircle[fillstyle=solid,fillcolor=fillval](1.695652,0.021739){0.021739}
\pscircle[fillstyle=solid,fillcolor=fillval](1.956522,0.021739){0.021739}
\pscircle[fillstyle=solid,fillcolor=fillval](2.217391,0.021739){0.021739}
\pscircle[fillstyle=solid,fillcolor=fillval](2.478261,0.021739){0.021739}
\psarcn(0.413043,0.021739){0.130435}{-180}{-360}
\psarcn(0.413043,0.021739){0.391304}{-180}{-360}
\psarcn(1.826087,0.043478){0.130435}{179.999998}{0.000002}
\psarcn(2.347826,0.043478){0.130435}{179.999998}{0.000002}
\rput(1.26087,0){$=$}
\uput{0.5ex}[d](0.021739,-0.021739){$A^*$}
\uput{0.5ex}[d](0.543478,-0.021739){$A^*$}
\uput{0.5ex}[d](1.695652,0){$A^*$}
\uput{0.5ex}[d](2.217391,0){$A^*$}
\uput{0.5ex}[d](0.282609,-0.021739){$A$}
\uput{0.5ex}[d](0.804348,-0.021739){$A$}
\uput{0.5ex}[d](1.956522,0){$A$}
\uput{0.5ex}[d](2.478261,0){$A$}
\end{pspicture}%
\end{center}

\noindent This amounts to a ``confusion of entanglements''.

In fact, we shall find it more convenient to prove this result in the following form:
\[ \eta_{A} \otimes \eta_{A} = (3\ 2\ 1\ 4) \circ (\eta_{A} \otimes \eta_{A})  \]
Here $(3\ 2\ 1\ 4)$ is the permutation acting on the tensor product of four factors which is built from the symmetry isomorphisms in the obvious fashion.
Diagrammatically:
\begin{center}
\psset{unit=1in,cornersize=absolute,dimen=middle}%
\begin{pspicture}(0,-0.568966)(3.025862,0.181034)%
% dpic version 08.Jul.05 for PSTricks 0.93a
\psset{linewidth=1pt}%
\newgray{fillval}{0.3}
\pscircle[fillstyle=solid,fillcolor=fillval](0.025862,0){0.025862}
\pscircle[fillstyle=solid,fillcolor=fillval](0.336207,0){0.025862}
\pscircle[fillstyle=solid,fillcolor=fillval](0.646552,0){0.025862}
\pscircle[fillstyle=solid,fillcolor=fillval](0.956897,0){0.025862}
\pscircle[fillstyle=solid,fillcolor=fillval](2.043103,0){0.025862}
\pscircle[fillstyle=solid,fillcolor=fillval](2.353448,0){0.025862}
\pscircle[fillstyle=solid,fillcolor=fillval](2.663793,0){0.025862}
\pscircle[fillstyle=solid,fillcolor=fillval](2.974138,0){0.025862}
\pscircle[fillstyle=solid,fillcolor=fillval](2.068966,-0.543103){0.025862}
\pscircle[fillstyle=solid,fillcolor=fillval](2.37931,-0.543103){0.025862}
\pscircle[fillstyle=solid,fillcolor=fillval](2.689655,-0.543103){0.025862}
\pscircle[fillstyle=solid,fillcolor=fillval](3,-0.543103){0.025862}
\psarcn(0.181034,0.025862){0.155172}{179.999999}{0.000001}
\psarcn(0.801724,0.025862){0.155172}{-180}{-360}
\psarcn(2.198276,0.025862){0.155172}{179.999998}{0.000002}
\psarcn(2.818966,0.025862){0.155172}{179.999998}{0.000002}
\psline(2.043103,-0.025862)(2.689655,-0.517241)
\psline(2.353448,-0.025862)(2.37931,-0.517241)
\psline(2.663793,-0.025862)(2.068966,-0.517241)
\psline(2.974138,-0.025862)(3,-0.517241)
\rput(1.5,0){$=$}
\end{pspicture}%
\end{center}

\begin{lemma}
\label{sidemlemm}
We have $\Delta_{\II} = l_{\II}^{-1} : \II \rarr \II \otimes \II$.
\end{lemma}
\begin{proof}
This is an immediate application of the monoidality of $\Delta$, together with $e = l_{\II}^{-1}$ for the codomain functor.
\end{proof}

\begin{lemma}
\label{pstep1}
Let $u : \II \rarr A \otimes B$ be a morphism in a symmetric monoidal category with cloning.
Then
\[ u \otimes u = (3\ 2\ 1\ 4) \circ (u \otimes u) \, . \]
\end{lemma}

\begin{proof} 
Consider the following diagram.
\[ \begin{diagram}
I & \rTo^{\Delta_{I}} & I \otimes I \\
\dTo^{u} & & \dTo_{u \otimes u} \\
A \otimes B & \rTo^{\Delta_{A \otimes B}} & A \otimes B \otimes A \otimes B \\
\dTo^{\Delta_{A} \otimes \Delta_{B}} & \rdTo_{\Delta_{A} \otimes \Delta_{B}} & \uTo_{1 \otimes \sigma \otimes 1} \\
A \otimes A \otimes B \otimes B & \rTo_{\sigma \otimes 1} & A \otimes A \otimes B \otimes B
\end{diagram}
\]
The upper square commutes by naturality of $\Delta$. The upper triangle of the lower square commutes by monoidality of $\Delta$. The lower triangle commutes by cocommutativity of $\Delta$ in the first component, and then tensoring with the second component and using the bifunctoriality of the tensor.

Let $f = (u \otimes u) \circ \Delta_{\II}$, and  $g = (\Delta_{A} \otimes \Delta_{B}) \circ u$. Then by the above diagram
\[ f = (1 \otimes \sigma \otimes 1) \circ (\sigma \otimes 1) \circ g \, . \]
A simple computation with permutations shows that
\[ (1 \otimes \sigma \otimes 1) \circ (\sigma \otimes 1) = (1\ 3\ 2\ 4) \circ (2\ 1\ 3\ 4) = (3\ 2\ 1\ 4) \circ (1 \otimes \sigma \otimes 1) \, . \]
Appealing to the above diagram again, $f = (1 \otimes \sigma \otimes 1) \circ g$. Hence
\[ \begin{array}{rcl}
f & = &  (1 \otimes \sigma \otimes 1) \circ (\sigma \otimes 1) \circ g \\
& = & (3\ 2\ 1\ 4) \circ (1 \otimes \sigma \otimes 1) \circ g \\
& = & (3\ 2\ 1\ 4) \circ f \, .
\end{array}
\]
Applying the previous lemma:
\[ u \otimes u = f \circ l_{\II} = (3\ 2\ 1\ 4) \circ f \circ l_{\II} = (3\ 2\ 1\ 4) \circ (u \otimes u) \, . \]
Diagrammatically, this can be presented as follows:
\begin{center}
\psset{unit=1in,cornersize=absolute,dimen=middle}%
\begin{pspicture}(0,-1.098901)(4,0.193407)%
% dpic version 28.Jun.06 for PSTricks 0.93a
\psset{linewidth=1pt}%
\newgray{fillval}{0.3}
\pscircle[fillstyle=solid,fillcolor=fillval](0.008791,0){0.008791}
\pscircle[fillstyle=solid,fillcolor=fillval](0.202198,0){0.008791}
\pscircle[fillstyle=solid,fillcolor=fillval](0.395604,0){0.008791}
\pscircle[fillstyle=solid,fillcolor=fillval](0.589011,0){0.008791}
\pscircle[fillstyle=solid,fillcolor=fillval](1.30989,0){0.008791}
\pscircle[fillstyle=solid,fillcolor=fillval](1.679121,0){0.008791}
\pscircle[fillstyle=solid,fillcolor=fillval](2.4,0){0.008791}
\pscircle[fillstyle=solid,fillcolor=fillval](2.769231,0){0.008791}
\pscircle[fillstyle=solid,fillcolor=fillval](3.49011,0){0.008791}
\pscircle[fillstyle=solid,fillcolor=fillval](3.859341,0){0.008791}
\pscircle[fillstyle=solid,fillcolor=fillval](1.230769,-0.36044){0.008791}
\pscircle[fillstyle=solid,fillcolor=fillval](1.424176,-0.36044){0.008791}
\pscircle[fillstyle=solid,fillcolor=fillval](1.6,-0.36044){0.008791}
\pscircle[fillstyle=solid,fillcolor=fillval](1.793407,-0.36044){0.008791}
\pscircle[fillstyle=solid,fillcolor=fillval](2.320879,-0.36044){0.008791}
\pscircle[fillstyle=solid,fillcolor=fillval](2.514286,-0.36044){0.008791}
\pscircle[fillstyle=solid,fillcolor=fillval](2.69011,-0.36044){0.008791}
\pscircle[fillstyle=solid,fillcolor=fillval](2.883516,-0.36044){0.008791}
\pscircle[fillstyle=solid,fillcolor=fillval](3.410989,-0.36044){0.008791}
\pscircle[fillstyle=solid,fillcolor=fillval](3.604396,-0.36044){0.008791}
\pscircle[fillstyle=solid,fillcolor=fillval](3.78022,-0.36044){0.008791}
\pscircle[fillstyle=solid,fillcolor=fillval](3.973626,-0.36044){0.008791}
\pscircle[fillstyle=solid,fillcolor=fillval](1.23956,-0.720879){0.008791}
\pscircle[fillstyle=solid,fillcolor=fillval](1.432967,-0.720879){0.008791}
\pscircle[fillstyle=solid,fillcolor=fillval](1.608791,-0.720879){0.008791}
\pscircle[fillstyle=solid,fillcolor=fillval](1.802198,-0.720879){0.008791}
\pscircle[fillstyle=solid,fillcolor=fillval](2.32967,-0.720879){0.008791}
\pscircle[fillstyle=solid,fillcolor=fillval](2.523077,-0.720879){0.008791}
\pscircle[fillstyle=solid,fillcolor=fillval](2.698901,-0.720879){0.008791}
\pscircle[fillstyle=solid,fillcolor=fillval](2.892308,-0.720879){0.008791}
\pscircle[fillstyle=solid,fillcolor=fillval](3.41978,-0.720879){0.008791}
\pscircle[fillstyle=solid,fillcolor=fillval](3.613187,-0.720879){0.008791}
\pscircle[fillstyle=solid,fillcolor=fillval](3.789011,-0.720879){0.008791}
\pscircle[fillstyle=solid,fillcolor=fillval](3.982418,-0.720879){0.008791}
\pscircle[fillstyle=solid,fillcolor=fillval](2.338462,-1.081319){0.008791}
\pscircle[fillstyle=solid,fillcolor=fillval](2.531868,-1.081319){0.008791}
\pscircle[fillstyle=solid,fillcolor=fillval](2.707692,-1.081319){0.008791}
\pscircle[fillstyle=solid,fillcolor=fillval](2.901099,-1.081319){0.008791}
\pscircle[fillstyle=solid,fillcolor=fillval](3.428571,-1.081319){0.008791}
\pscircle[fillstyle=solid,fillcolor=fillval](3.621978,-1.081319){0.008791}
\pscircle[fillstyle=solid,fillcolor=fillval](3.797802,-1.081319){0.008791}
\pscircle[fillstyle=solid,fillcolor=fillval](3.991209,-1.081319){0.008791}
\psarcn(0.105495,0.008791){0.096703}{-180}{-360}
\psarcn(0.492308,0.008791){0.096703}{-180}{-360}
\psarcn(1.494505,0.008791){0.184615}{-180}{-360}
\psarcn(2.584615,0.008791){0.184615}{-180}{-360}
\psarcn(3.674725,0.008791){0.184615}{-180}{-360}
\psline(1.30989,-0.008791)(1.230769,-0.351648)
\psline(1.30989,-0.008791)(1.424176,-0.351648)
\psline(1.679121,-0.008791)(1.6,-0.351648)
\psline(1.679121,-0.008791)(1.793407,-0.351648)
\psline(2.4,-0.008791)(2.320879,-0.351648)
\psline(2.4,-0.008791)(2.514286,-0.351648)
\psline(2.769231,-0.008791)(2.69011,-0.351648)
\psline(2.769231,-0.008791)(2.883516,-0.351648)
\psline(3.49011,-0.008791)(3.410989,-0.351648)
\psline(3.49011,-0.008791)(3.604396,-0.351648)
\psline(3.859341,-0.008791)(3.78022,-0.351648)
\psline(3.859341,-0.008791)(3.973626,-0.351648)
\psline(1.230769,-0.369231)(1.23956,-0.712088)
\psline(1.424176,-0.369231)(1.608791,-0.712088)
\psline(1.6,-0.369231)(1.432967,-0.712088)
\psline(1.793407,-0.369231)(1.802198,-0.712088)
\psline(2.320879,-0.369231)(2.523077,-0.712088)
\psline(2.514286,-0.369231)(2.32967,-0.712088)
\psline(2.69011,-0.369231)(2.698901,-0.712088)
\psline(2.883516,-0.369231)(2.892308,-0.712088)
\psline(3.410989,-0.369231)(3.41978,-0.712088)
\psline(3.604396,-0.369231)(3.789011,-0.712088)
\psline(3.78022,-0.369231)(3.613187,-0.712088)
\psline(3.973626,-0.369231)(3.982418,-0.712088)
\pscircle[fillstyle=solid,fillcolor=fillval](1.23956,-1.09011){0.008791}
\pscircle[fillstyle=solid,fillcolor=fillval](1.432967,-1.09011){0.008791}
\pscircle[fillstyle=solid,fillcolor=fillval](1.608791,-1.09011){0.008791}
\pscircle[fillstyle=solid,fillcolor=fillval](1.802198,-1.09011){0.008791}
\psline(1.23956,-0.72967)(1.23956,-1.081319)
\psline(1.432967,-0.72967)(1.432967,-1.081319)
\psline(1.608791,-0.72967)(1.608791,-1.081319)
\psline(1.802198,-0.72967)(1.802198,-1.081319)
\psline(2.32967,-0.72967)(2.338462,-1.072527)
\psline(2.523077,-0.72967)(2.707692,-1.072527)
\psline(2.698901,-0.72967)(2.531868,-1.072527)
\psline(2.892308,-0.72967)(2.901099,-1.072527)
\psline(3.41978,-0.72967)(3.797802,-1.072527)
\psline(3.613187,-0.72967)(3.621978,-1.072527)
\psline(3.789011,-0.72967)(3.428571,-1.072527)
\psline(3.982418,-0.72967)(3.991209,-1.072527)
\rput(0.949451,0){$=$}
\rput(2.03956,0){$=$}
\rput(3.12967,0){$=$}
\end{pspicture}%
\end{center}
and hence
\begin{center}
\psset{unit=1in,cornersize=absolute,dimen=middle}%
\begin{pspicture}(0,-0.568966)(3.025862,0.181034)%
% dpic version 08.Jul.05 for PSTricks 0.93a
\psset{linewidth=1pt}%
\newgray{fillval}{0.3}
\pscircle[fillstyle=solid,fillcolor=fillval](0.025862,0){0.025862}
\pscircle[fillstyle=solid,fillcolor=fillval](0.336207,0){0.025862}
\pscircle[fillstyle=solid,fillcolor=fillval](0.646552,0){0.025862}
\pscircle[fillstyle=solid,fillcolor=fillval](0.956897,0){0.025862}
\pscircle[fillstyle=solid,fillcolor=fillval](2.043103,0){0.025862}
\pscircle[fillstyle=solid,fillcolor=fillval](2.353448,0){0.025862}
\pscircle[fillstyle=solid,fillcolor=fillval](2.663793,0){0.025862}
\pscircle[fillstyle=solid,fillcolor=fillval](2.974138,0){0.025862}
\pscircle[fillstyle=solid,fillcolor=fillval](2.068966,-0.543103){0.025862}
\pscircle[fillstyle=solid,fillcolor=fillval](2.37931,-0.543103){0.025862}
\pscircle[fillstyle=solid,fillcolor=fillval](2.689655,-0.543103){0.025862}
\pscircle[fillstyle=solid,fillcolor=fillval](3,-0.543103){0.025862}
\psarcn(0.181034,0.025862){0.155172}{179.999999}{0.000001}
\psarcn(0.801724,0.025862){0.155172}{-180}{-360}
\psarcn(2.198276,0.025862){0.155172}{179.999998}{0.000002}
\psarcn(2.818966,0.025862){0.155172}{179.999998}{0.000002}
\psline(2.043103,-0.025862)(2.689655,-0.517241)
\psline(2.353448,-0.025862)(2.37931,-0.517241)
\psline(2.663793,-0.025862)(2.068966,-0.517241)
\psline(2.974138,-0.025862)(3,-0.517241)
\rput(1.5,0){$=$}
\end{pspicture}%
\end{center}
\end{proof}

\noindent Note that this lemma is proved in generality, for any morphism $u$ of the required shape. However, we shall, as expected, apply it by taking $u = \eta_{A}$. It will be convenient to give the remainder of the proof in diagrammatic form.

%\subsection{Proving the cloning collapse theorem ctd}
\paragraph{Second step}
We use the first step to show that
\begin{center}
\fbox{\textbf{the twist map $=$ the identity}}
\end{center}
in a compact category with cloning,
by putting parallel and serial caps in a common context and simplifying using the triangular identities.

The context is:
\begin{center}
\psset{unit=1in,cornersize=absolute,dimen=middle}%
\begin{pspicture}(-0,-0.987808)(1.340372,0.512192)%
% dpic version 08.Jul.05 for PSTricks 0.93a
\psset{linewidth=1pt}%
\newgray{fillval}{0.3}
\pscircle[fillstyle=solid,fillcolor=fillval](0.021619,0){0.021619}
\pscircle[fillstyle=solid,fillcolor=fillval](0.281046,0){0.021619}
\pscircle[fillstyle=solid,fillcolor=fillval](0.540473,0){0.021619}
\pscircle[fillstyle=solid,fillcolor=fillval](0.7999,0){0.021619}
\pscircle[fillstyle=solid,fillcolor=fillval](1.059327,0){0.021619}
\pscircle[fillstyle=solid,fillcolor=fillval](0.7999,-0.475616){0.021619}
\pscircle[fillstyle=solid,fillcolor=fillval](1.059327,-0.475616){0.021619}
\pscircle[fillstyle=solid,fillcolor=fillval](1.318754,-0.475616){0.021619}
\pscircle[fillstyle=solid,fillcolor=fillval](0.021619,0.475616){0.021619}
\pscircle[fillstyle=solid,fillcolor=fillval](1.318754,0.475616){0.021619}
\pscircle[fillstyle=solid,fillcolor=fillval](0.540473,-0.951232){0.021619}
\pscircle[fillstyle=solid,fillcolor=fillval](0.7999,-0.951232){0.021619}
\psarc(0.151332,-0.021619){0.129713}{-179.999999}{-0.000001}
\psarc(1.18904,-0.497235){0.129713}{-179.999998}{-0.000002}
\psline(0.021619,0.021619)(0.021619,0.453997)
\psline(1.318754,-0.453997)(1.318754,0.453997)
\psline(0.540473,-0.021619)(0.540473,-0.929613)
\psline(0.540473,-0.497235)(0.540473,-0.929613)
\psline(0.815187,-0.015287)(1.04404,-0.460329)
\psline(1.04404,-0.015287)(0.815187,-0.460329)
\psline(0.7999,-0.497235)(0.7999,-0.929613)
\uput{0.5ex}[u](0.021619,0.497235){$A$}
\uput{0.5ex}[u](1.318754,0.497235){$A$}
\uput{0.5ex}[d](0.540473,-0.972851){$A$}
\uput{0.5ex}[d](0.7999,-0.972851){$A$}
\uput{0.5ex}[u](0.281046,0.021619){$A^*$}
\uput{0.5ex}[u](0.7999,0.021619){$A^*$}
\uput{0.5ex}[u](0.540473,0.021619){$A$}
\uput{0.5ex}[u](1.059327,0.021619){$A$}
\end{pspicture}%
\end{center}

%\subsection{Identity = Twist}
\noindent We get:

\vspace{.5in}
\begin{center}
\psset{unit=1in,cornersize=absolute,dimen=middle}%
\begin{pspicture}(-0,-0.620813)(2.5,0.3219)%
% dpic version 08.Jul.05 for PSTricks 0.93a
\psset{linewidth=1pt}%
\newgray{fillval}{0.3}
\pscircle[fillstyle=solid,fillcolor=fillval](0.013587,0){0.013587}
\pscircle[fillstyle=solid,fillcolor=fillval](0.17663,0){0.013587}
\pscircle[fillstyle=solid,fillcolor=fillval](0.339674,0){0.013587}
\pscircle[fillstyle=solid,fillcolor=fillval](0.502717,0){0.013587}
\pscircle[fillstyle=solid,fillcolor=fillval](0.665761,0){0.013587}
\pscircle[fillstyle=solid,fillcolor=fillval](0.502717,-0.298913){0.013587}
\pscircle[fillstyle=solid,fillcolor=fillval](0.665761,-0.298913){0.013587}
\pscircle[fillstyle=solid,fillcolor=fillval](0.828804,-0.298913){0.013587}
\pscircle[fillstyle=solid,fillcolor=fillval](0.013587,0.298913){0.013587}
\pscircle[fillstyle=solid,fillcolor=fillval](0.828804,0.298913){0.013587}
\pscircle[fillstyle=solid,fillcolor=fillval](0.339674,-0.597826){0.013587}
\pscircle[fillstyle=solid,fillcolor=fillval](0.502717,-0.597826){0.013587}
\psarc(0.095109,-0.013587){0.081522}{-179.999999}{-0.000001}
\psarc(0.747283,-0.3125){0.081522}{-179.999998}{-0.000002}
\psline(0.013587,0.013587)(0.013587,0.285326)
\psline(0.828804,-0.285326)(0.828804,0.285326)
\psline(0.339674,-0.013587)(0.339674,-0.584239)
\psline(0.339674,-0.3125)(0.339674,-0.584239)
\psline(0.512325,-0.009607)(0.656153,-0.289306)
\psline(0.656153,-0.009607)(0.512325,-0.289306)
\psline(0.502717,-0.3125)(0.502717,-0.584239)
\psarcn(0.421196,0.013587){0.081522}{-180}{-360}
\psarcn(0.421196,0.013587){0.244565}{-180}{-360}
\pscircle[fillstyle=solid,fillcolor=fillval](1.929348,0.298913){0.013587}
\pscircle[fillstyle=solid,fillcolor=fillval](2.486413,0.298913){0.013587}
\pscircle[fillstyle=solid,fillcolor=fillval](1.929348,-0.584239){0.013587}
\pscircle[fillstyle=solid,fillcolor=fillval](2.486413,-0.584239){0.013587}
\psline(1.938955,0.289306)(2.476806,-0.574632)
\psline(2.476806,0.289306)(1.938955,-0.574632)
\rput(1.38587,-0.298913){$=$}
\uput{0.5ex}[u](0.013587,0.3125){$A$}
\uput{0.5ex}[u](0.828804,0.3125){$A$}
\uput{0.5ex}[d](0.339674,-0.611413){$A$}
\uput{0.5ex}[d](0.502717,-0.611413){$A$}
\uput{0.5ex}[u](1.929348,0.3125){$A$}
\uput{0.5ex}[u](2.486413,0.3125){$A$}
\uput{0.5ex}[d](1.929348,-0.597826){$A$}
\uput{0.5ex}[d](2.486413,-0.597826){$A$}
\end{pspicture}%
\end{center}

and:
\vspace{.1in}
\begin{center}
\psset{unit=1in,cornersize=absolute,dimen=middle}%
\begin{pspicture}(-0,-0.620813)(2.5,0.3219)%
% dpic version 08.Jul.05 for PSTricks 0.93a
\psset{linewidth=1pt}%
\newgray{fillval}{0.3}
\pscircle[fillstyle=solid,fillcolor=fillval](0.013587,0){0.013587}
\pscircle[fillstyle=solid,fillcolor=fillval](0.17663,0){0.013587}
\pscircle[fillstyle=solid,fillcolor=fillval](0.339674,0){0.013587}
\pscircle[fillstyle=solid,fillcolor=fillval](0.502717,0){0.013587}
\pscircle[fillstyle=solid,fillcolor=fillval](0.665761,0){0.013587}
\pscircle[fillstyle=solid,fillcolor=fillval](0.502717,-0.298913){0.013587}
\pscircle[fillstyle=solid,fillcolor=fillval](0.665761,-0.298913){0.013587}
\pscircle[fillstyle=solid,fillcolor=fillval](0.828804,-0.298913){0.013587}
\pscircle[fillstyle=solid,fillcolor=fillval](0.013587,0.298913){0.013587}
\pscircle[fillstyle=solid,fillcolor=fillval](0.828804,0.298913){0.013587}
\pscircle[fillstyle=solid,fillcolor=fillval](0.339674,-0.597826){0.013587}
\pscircle[fillstyle=solid,fillcolor=fillval](0.502717,-0.597826){0.013587}
\psarc(0.095109,-0.013587){0.081522}{-179.999999}{-0.000001}
\psarc(0.747283,-0.3125){0.081522}{-179.999998}{-0.000002}
\psline(0.013587,0.013587)(0.013587,0.285326)
\psline(0.828804,-0.285326)(0.828804,0.285326)
\psline(0.339674,-0.013587)(0.339674,-0.584239)
\psline(0.339674,-0.3125)(0.339674,-0.584239)
\psline(0.512325,-0.009607)(0.656153,-0.289306)
\psline(0.656153,-0.009607)(0.512325,-0.289306)
\psline(0.502717,-0.3125)(0.502717,-0.584239)
\psarcn(0.584239,0.013587){0.081522}{179.999999}{0.000001}
\psarcn(0.258152,0.013587){0.081522}{-180}{-360}
\pscircle[fillstyle=solid,fillcolor=fillval](1.929348,0.298913){0.013587}
\pscircle[fillstyle=solid,fillcolor=fillval](2.486413,0.298913){0.013587}
\pscircle[fillstyle=solid,fillcolor=fillval](1.929348,-0.584239){0.013587}
\pscircle[fillstyle=solid,fillcolor=fillval](2.486413,-0.584239){0.013587}
\psline(1.929348,0.285326)(1.929348,-0.570652)
\psline(2.486413,0.285326)(2.486413,-0.570652)
\rput(1.38587,-0.298913){$=$}
\uput{0.5ex}[u](0.013587,0.3125){$A$}
\uput{0.5ex}[u](0.828804,0.3125){$A$}
\uput{0.5ex}[d](0.339674,-0.611413){$A$}
\uput{0.5ex}[d](0.502717,-0.611413){$A$}
\uput{0.5ex}[u](1.929348,0.3125){$A$}
\uput{0.5ex}[u](2.486413,0.3125){$A$}
\uput{0.5ex}[d](1.929348,-0.597826){$A$}
\uput{0.5ex}[d](2.486413,-0.597826){$A$}
\end{pspicture}%
\end{center}

\noindent We used the original picture of nested caps for clarity. If we use the picture directly corresponding to the statement of lemma~\ref{pstep1}, we obtain the same result:
\vspace{.1in}
\begin{center}
\psset{unit=1in,cornersize=absolute,dimen=middle}%
\begin{pspicture}(-0,-0.744976)(3,0.472826)%
% dpic version 08.Jul.05 for PSTricks 0.93a
\psset{linewidth=1pt}%
\newgray{fillval}{0.3}
\pscircle[fillstyle=solid,fillcolor=fillval](0.016304,0){0.016304}
\pscircle[fillstyle=solid,fillcolor=fillval](0.211957,0){0.016304}
\pscircle[fillstyle=solid,fillcolor=fillval](0.407609,0){0.016304}
\pscircle[fillstyle=solid,fillcolor=fillval](0.603261,0){0.016304}
\pscircle[fillstyle=solid,fillcolor=fillval](0.798913,0){0.016304}
\pscircle[fillstyle=solid,fillcolor=fillval](0.603261,-0.358696){0.016304}
\pscircle[fillstyle=solid,fillcolor=fillval](0.798913,-0.358696){0.016304}
\pscircle[fillstyle=solid,fillcolor=fillval](0.994565,-0.358696){0.016304}
\pscircle[fillstyle=solid,fillcolor=fillval](0.016304,0.358696){0.016304}
\pscircle[fillstyle=solid,fillcolor=fillval](0.211957,0.358696){0.016304}
\pscircle[fillstyle=solid,fillcolor=fillval](0.407609,0.358696){0.016304}
\pscircle[fillstyle=solid,fillcolor=fillval](0.603261,0.358696){0.016304}
\pscircle[fillstyle=solid,fillcolor=fillval](0.798913,0.358696){0.016304}
\pscircle[fillstyle=solid,fillcolor=fillval](0.994565,0.358696){0.016304}
\pscircle[fillstyle=solid,fillcolor=fillval](0.407609,-0.717391){0.016304}
\pscircle[fillstyle=solid,fillcolor=fillval](0.603261,-0.717391){0.016304}
\psarc(0.11413,-0.016304){0.097826}{-179.999999}{-0.000001}
\psarc(0.896739,-0.375){0.097826}{-179.999998}{-0.000002}
\psline(0.016304,0.342391)(0.016304,0.016304)
\psline(0.798913,0.342391)(0.798913,0.016304)
\psline(0.994565,0.342391)(0.994565,-0.342391)
\psline(0.603261,0.016304)(0.211957,0.342391)
\psline(0.994565,-0.342391)(0.994565,0.342391)
\psline(0.407609,0.342391)(0.407609,0.016304)
\psline(0.603261,0.342391)(0.211957,0.016304)
\psline(0.407609,-0.016304)(0.407609,-0.701087)
\psline(0.407609,-0.375)(0.407609,-0.701087)
\psline(0.61479,-0.011529)(0.787384,-0.347167)
\psline(0.787384,-0.011529)(0.61479,-0.347167)
\psline(0.603261,-0.375)(0.603261,-0.701087)
\psarcn(0.309783,0.375){0.097826}{180}{0}
\psarcn(0.701087,0.375){0.097826}{180}{0}
\pscircle[fillstyle=solid,fillcolor=fillval](2.315217,0.358696){0.016304}
\pscircle[fillstyle=solid,fillcolor=fillval](2.983696,0.358696){0.016304}
\pscircle[fillstyle=solid,fillcolor=fillval](2.315217,-0.701087){0.016304}
\pscircle[fillstyle=solid,fillcolor=fillval](2.983696,-0.701087){0.016304}
\psline(2.326746,0.347167)(2.972167,-0.689558)
\psline(2.972167,0.347167)(2.326746,-0.689558)
\rput(1.663043,-0.358696){$=$}
\uput{0.5ex}[u](0.016304,0.375){$A$}
\uput{0.5ex}[u](0.994565,0.375){$A$}
\uput{0.5ex}[d](0.407609,-0.733696){$A$}
\uput{0.5ex}[d](0.603261,-0.733696){$A$}
\uput{0.5ex}[u](2.315217,0.375){$A$}
\uput{0.5ex}[u](2.983696,0.375){$A$}
\uput{0.5ex}[d](2.315217,-0.717391){$A$}
\uput{0.5ex}[d](2.983696,-0.717391){$A$}
\end{pspicture}%
\end{center}
The important point is that the left input is connected to the right  output, and the right input to the left output.

%\subsection{The \textit{coup de gr\^{a}ce}}

\paragraph{Third step}
Finally, we use the trace to show that any endomorphism $f : A \lrarr A$ is \emph{a scalar multiple of the identity}:
\[ f = s \sdot 1_{A} \]
for $s = \Tr(f)$.

\begin{center}
\psset{unit=1in,cornersize=absolute,dimen=middle}%
\begin{pspicture}(0,-0.893939)(3.606061,0.106061)%
% dpic version 08.Jul.05 for PSTricks 0.93a
\psset{linewidth=1pt}%
\newgray{fillval}{0.3}
\pscircle[fillstyle=solid,fillcolor=fillval](0.015152,0){0.015152}
\pscircle[fillstyle=solid,fillcolor=fillval](0.015152,-0.787879){0.015152}
\psframe(0.272727,-0.469697)(0.424242,-0.318182)
\rput(0.348485,-0.393939){$f$}
\psline(0.015152,-0.015152)(0.015152,-0.772727)
\psline(0.348485,0.015152)(0.348485,-0.318182)
\psline(0.348485,-0.469697)(0.348485,-0.80303)
\psline(0.530303,0.015152)(0.530303,-0.80303)
\psarcn(0.439394,0.015152){0.090909}{-180}{-360}
\psarc(0.439394,-0.80303){0.090909}{180}{360}
\uput{0.5ex}[u](0.015152,0.015152){$A$}
\uput{0.5ex}[d](0.015152,-0.80303){$A$}
\rput(1.181818,-0.393939){$=$}
\pscircle[fillstyle=solid,fillcolor=fillval](1.772727,0){0.015152}
\pscircle[fillstyle=solid,fillcolor=fillval](1.772727,-0.787879){0.015152}
\psframe(2.030303,-0.469697)(2.181818,-0.318182)
\rput(2.106061,-0.393939){$f$}
\psline(1.783441,-0.010714)(2.106061,-0.166667)
(2.106061,-0.318182)
\psline(2.106061,0.015152)(1.772727,-0.166667)
(1.772727,-0.772727)
\psline(2.106061,-0.469697)(2.106061,-0.80303)
\psline(2.287879,0.015152)(2.287879,-0.80303)
\psarcn(2.19697,0.015152){0.090909}{179.999998}{0.000002}
\psarc(2.19697,-0.80303){0.090909}{-179.999998}{-0.000002}
\uput{0.5ex}[u](1.772727,0.015152){$A$}
\uput{0.5ex}[d](1.772727,-0.80303){$A$}
\rput(2.939394,-0.393939){$=$}
\pscircle[fillstyle=solid,fillcolor=fillval](3.530303,0){0.015152}
\pscircle[fillstyle=solid,fillcolor=fillval](3.530303,-0.787879){0.015152}
\psframe(3.454545,-0.469697)(3.606061,-0.318182)
\rput(3.530303,-0.393939){$f$}
\psline(3.530303,-0.015152)(3.530303,-0.318182)
\psline(3.530303,-0.469697)(3.530303,-0.80303)
\uput{0.5ex}[u](3.530303,0.015152){$A$}
\uput{0.5ex}[d](3.530303,-0.80303){$A$}
\end{pspicture}%
\end{center}

\noindent This completes the proof of the Cloning Collapse Theorem~\ref{thethe}.

\subsection{Examples}

We note another consequence of cloning.

\begin{proposition}
In a monoidal category with cloning, the multiplication of scalars is idempotent.
\end{proposition}

\begin{proof}
This follows immediately from naturality
\[ \begin{diagram}
I & \rTo^{\Delta_{I}} & I \otimes I \\
\dTo^{s} & & \dTo_{s \otimes s} \\
I & \rTo_{\Delta_{I}} & I \otimes I
\end{diagram}
\]
together with lemma~\ref{sidemlemm}.
\end{proof}
Thus the scalars form a commutative, idempotent monoid, \ie a semilattice.

Given any semilattice $S$, we regard it qua monoid as a one-object category, say with object $\Obj$. We can define a trivial strict monoidal structure on this category, with
\[ \Obj \otimes \Obj = \Obj = \II \, . \]
Bifunctoriality follows from commutativity. A natural diagonal is also given trivially by the identity element (which is the top element of the induced partial order, if we view the semilattice operation as meet). Units and counits are also given trivially by the identity. Note that the scalars in this category are of course just the elements of $S$.

Thus any semilattice yields an example of a (trivial) compact category with cloning. Note the contrast with Joyal's lemma. While every boolean algebra is of course a semilattice, it forms a degenerate cartesian closed category as a \emph{poset}, with many objects but at most one morphism in each homset. The degenerate categories we are considering are categories qua \emph{monoids}, with arbitrarily large hom-sets, but only one object. Posets and monoids are opposite extremal examples of categories, which appear as  contrasting degenerate examples allowed by these no-go results.

Note that our result as it stands is not directly comparable with Joyal's, since our hypotheses are weaker insofar as we only assume a monoidal diagonal rather than full cartesian structure, but stronger insofar as we assume compact closure. A boolean algebra which is compact closed qua category is necessarily the trivial, one-element poset, since meets and joins --- and in particular the top and bottom of the lattice --- are identified.

\subsection{Discussion}

The Cloning Collapse theorem can be read as a No-Go theorem. It says that it is not possible to combine basic structural features of quantum entanglement with a uniform cloning operation without collapsing to degeneracy. It should be understood that the key point here is the \emph{uniformity} of the cloning operation, which is formalized as the \emph{monoidal naturality} of the diagonal. A suitable intuition is to think of this as corresponding to \emph{basis-independence}.\footnote{In fact, the original example which led Eilenberg and Mac Lane to define naturality was the naturality of the isomorphism from a finite-dimensional vector space to its second dual, as compared with the non-natural isomorphism to the first dual.} The distinction is between an operation that exists in a representation-independent form, for logical reasons, as compared to operations which do intrinsically depend on specific representations.

In fact, in turns out that much significant quantum structure can be captured in our categorical setting by \emph{non-uniform} copying operations \cite{CPav}. Given a choice of basis for a finite-dimensional Hilbert space $\HH$, one can define a diagonal
\[ \ket{i} \mapsto \ket{ii} \, .  \]
This is coassociative and cocommutative, and extends to a comonoid structure. Applying the dagger yields a commutative monoid structures, and the two structures interact by  the Frobenius law. It can be shown that such ``dagger Frobenius structures'' on finite-dimensional Hilbert spaces correspond exactly to bases. Since bases correspond to ``choice of measurement context'', these structures can be used to formalize quantum measurements, and quantum protocols involving such measurements \cite{CPav}.

It is of the essence of quantum mechanics that \emph{many} such structures can coexist on the same system, leading to the idea of \emph{incompatible measurements}. This too has been axiomatized in the categorical setting, enabling  the effective description of many central features of quantum computation \cite{CD}.

Thus the No-Go result is delicately poised on the issue of naturality. It seems possible that a rather sharp delineation between quantum and classical, and more generally a classification of the space of possible theories incorporating various features, may be achieved by further development of these ideas.

\section{No-Deleting}
The issue of No-deleting is much simpler from the current perspective. A uniform deleting operation is a monoidal natural transformation $d_{A} : A \rarr \II$. Note that the domain of this transformation is the identity functor, while the codomain is the constant functor valued at $\II$.
The following result was originally observed by Bob Coecke in the dagger compact case:

\begin{proposition}
If a compact category has uniform deleting, then it is a preorder.
\end{proposition}

\begin{proof}
Given $f : A \lrarr B$, consider the naturality square
\[  \begin{diagram}
A \otimes B^{*} & \rTo^{d_{A \otimes B^{*}}} & \II \\
\dTo^{\coname{f}} &  & \deq \\
\II & \rTo_{d_{I}} & \II
\end{diagram}
\]
By monoidal naturality, $d_{I} = 1_{I}$. So for all $f, g : A \lrarr B$:
\[ \coname{f} = d_{A \otimes B^{*}} = \coname{g} \]
and hence $f = g$.
\end{proof}

\section{Further Directions}

We conclude by discussing some further developments and possible extensions of these ideas.

\begin{itemize}
\item In a forthcoming joint paper with Bob Coecke, the results are extended to cover No-Broadcasting by lifting the Cloning Collapse theorem to the CPM category \cite{Selinger}, which provides a categorical treatment of \emph{mixed states}.

\item The proof of the Cloning Collapse theorem makes essential use of  compactness. \\
\textbf{Open Question} Are there non-trivial examples of \emph{$*$-autonomous categories with uniform cloning operations}? \\
One can also consider various possible sharpenings of our results, by weakening the hypotheses, e.g. on monoidality of the diagonal, or by strengthening the conclusions, to a more definitive form of collapse.

\item Finally, the r\^ole of scalars in these results hints at the relevance of projectivity ideas \cite{deLL}, which should be developed further in the abstract setting.
\end{itemize}

\noindent Altogether, these results, while preliminary, suggest that the categorical axiomatization of quantum mechanics in \cite{AC2,AC3,AC4} does indeed open up a novel and fruitful perspective on No-Go Theorems and other foundational results. Moreover, these foundational topics in physics can usefully be informed by results and concepts stemming from categorical logic and theoretical computer science.

\bibliographystyle{alpha}

\end{document}